\documentclass{LMCS}
\usepackage{amssymb}
\usepackage{stmaryrd}
\usepackage{xspace}
\usepackage{enumerate}
\usepackage{pstricks,pst-node,pst-tree}
\usepackage{enumerate,hyperref}

\newfont{\fsc}{eusm10 scaled 1100}      % frenchscript letters
\newfont{\bbb}{bbm10 scaled 1100}       % blackboard bold
\newfont{\bbbs}{bbm10 scaled 800}       % blackboard bold scriptstyle

\newcommand{\inp}{\mathop\in}                           % takes less space
\newcommand{\Cdot}{\mathord\cdot}
\newcommand{\Defs}{\mathrel{:=}}
\newcommand{\Minus}{\mathord-}
\newcommand{\Eq}{\mathord=}

\newcommand{\CClose}{\mathord{\updownarrow}}
\newcommand{\IN}{\mathrel{\overline{\in}}}
\newcommand{\outc}{o}                                   % outcome
\newcommand{\Exp}[1]{\textrm{\rm Exp}_{#1}}
\renewcommand{\phi}{\varphi}
\newcommand{\aRel}{\ensuremath{\mathrel{{\mathcal R }}}}
\newcommand{\aRelS}{\ensuremath{\mathrel{{\mathcal S }}}}
\newcommand\Wide[1] {~~~#1~~~}
\newcommand{\plat}[1]{\raisebox{0pt}[0pt][0pt]{#1}}     % no vertical space

\newcommand{\BNFsep}{\;\;|\;\;}
\newcommand{\intc}{\mathrel{\sqcap}}
\newcommand{\extc}{\mathrel{\Box}}
\newcommand{\probc}[1]{\mathrel{\!_{\scriptscriptstyle #1}\oplus}}
\newcommand{\Par}[1]{\mathbin{|_{#1}}}
\newcommand{\parT}{\Par{\Act}}
\newcommand{\Cnil}{\mathop{\textbf{\textsf{0}}}}

\newcommand{\Act}{\ensuremath{\mathsf{Act}}\xspace}
\newcommand{\pCSP}{\ensuremath{\mathsf{pCSP}}\xspace}
\newcommand{\sCSP}{\ensuremath{\mathsf{sCSP}}\xspace}
\newcommand{\pCSPw}{\ensuremath{\mathsf{pCSP}^\eureka}\xspace}
\newcommand{\sCSPw}{\ensuremath{{\mathsf{sCSP}^\eureka}}\xspace}
\newcommand{\pCSPO}{\ensuremath{{\mathsf{pCSP}^\Omega}}\xspace}
\newcommand{\sCSPO}{\ensuremath{{\mathsf{sCSP}^\Omega}}\xspace}
\newcommand{\CSP}{\ensuremath{\mathsf{CSP}}\xspace}
\newcommand{\Op}[1]{\mbox{\bbb [}#1\mbox{\bbb ]}}
\newcommand{\Ops}[1]{\mbox{\bbbs [}#1\mbox{\bbbs ]}}

\newcommand{\DeltaC}{\Delta\kern -.17em^\circ}
\newcommand{\VV}{{\mathbb V}}
\newcommand{\Overline}[1]{\widehat{#1}}
\newcommand{\Apply}{\ensuremath{{\mathcal A}}\xspace}

\newcommand{\MDApply}{\ensuremath{{\Overline{\mathcal A}^{\raisebox{-.1em}{\scriptsize$\Omega$}}\xspace}}}

\newcommand{\MDCApply}{\ensuremath{{\Overline{\mathcal A}_{\scriptscriptstyle\updownarrow}^{\raisebox{-.1em}{\scriptsize$\Omega$}}\xspace}}}
\newcommand{\results}[1]{\ensuremath\VV(#1)}
\newcommand{\resultsN}{\ensuremath\VV\xspace}
\newcommand{\Dresults}[1]{\ensuremath{{\Overline\VV}}(#1)}
\newcommand{\DresultsN}{\ensuremath{{\Overline\VV}}\xspace}

\newcommand{\CDresults}[1]{\plat{$\Overline\VV\kern -0.2em_{\scriptscriptstyle\updownarrow}$}(#1)}
\newcommand{\CDresultsN}{\plat{$\Overline\VV\kern-0.2em_{\scriptscriptstyle\updownarrow}$}}
\newcommand{\MDCresults}[1]{\ensuremath{{\Overline\VV_{\scriptscriptstyle\updownarrow}^{\raisebox{-.3em}{\scriptsize$\Omega$}}(#1)}}}
\newcommand{\MDCresultsN}{\ensuremath{{\Overline\VV_{\scriptscriptstyle\updownarrow}^{\raisebox{-.3em}{\scriptsize$\Omega$}}\xspace}}}
\newcommand{\MDresults}[1]{\ensuremath{{\Overline\VV^\Omega(#1)}}}
\newcommand{\MDresultsN}{\ensuremath{{\Overline\VV^\Omega\xspace}}}
\newcommand{\subs}[1]{\mathop{\mbox{\fsc P}^{\mbox{\tiny\kern .1em$+$}}\kern-.15em(#1)}}
\newcommand{\subsN}{\mathop{\mbox{\fsc P}^{\mbox{\tiny\kern .1em$+$}}\kern-.15em}}
\newcommand{\eureka}{\ensuremath{\omega}\xspace}

\newcommand{\Hleq}{\mathrel{\leq_{\rm Ho}}}
\newcommand{\Sleq}{\mathrel{\leq_{\rm Sm}}}
\newcommand{\Mustleq}{\sqsubseteq_{\textrm{\rm must}}}
\newcommand{\Mayleq}{\sqsubseteq_{\textrm{\rm may}}}

\newcommand{\pMustleq}{\mathrel{\relax{\sqsubseteq}_{\textrm{\rm pmust}}}}
\newcommand{\pMayleq}{\mathrel{\relax{\sqsubseteq}_{\textrm{\rm pmay}}}}

\newcommand{\pMayeq}{\simeq_{\textrm{\rm pmay}}}
\newcommand{\dMustleq}{\mathrel{\Overline{\sqsubseteq}_{\textrm{\rm must}}}}
\newcommand{\dMayleq}{\mathrel{\Overline{\sqsubseteq}_{\textrm{\rm may}}}}
\newcommand{\pdMustleq}{\mathrel{\Overline{\sqsubseteq}_{\textrm{\rm pmust}}}}
\newcommand{\pdMayleq}{\mathrel{\Overline{\sqsubseteq}_{\textrm{\rm pmay}}}}
\newcommand{\pdmMustleq}{\mathrel{\Overline{\sqsubseteq}_{\textrm{\rm pmust}}^{\raisebox{-.3em}{\scriptsize$\Omega$}}}}
\newcommand{\pdmMayleq}{\mathrel{\Overline{\sqsubseteq}_{\textrm{\rm pmay}}^{\raisebox{-.3em}{\scriptsize$\Omega$}}}}

\newcommand{\forsim}{\mathrel{\lhd_{\raisebox{-.1em}{\tiny\it S}}}}
\newcommand{\forsiminv}{\mathrel{\rhd\kern -.3em_{\raisebox{-.34em}{\tiny\it S}}}}
\newcommand{\failsim}{\mathrel{\lhd_{\raisebox{-.1em}{\tiny\it FS}}}}
\newcommand{\failsime}{\mathrel{\lhd^e_{\raisebox{-.1em}{\tiny\it FS}}}}
\newcommand{\failsiminv}{\mathrel{\rhd\kern -.3em_{\raisebox{-.34em}{\tiny \it FS}}}}
\newcommand{\liftforsim}{\mathrel{\lift{\lhd_{\raisebox{-.1em}{\tiny\it S}}}}}
\newcommand{\liftfailsim}{\mathrel{\lift{\lhd_{\raisebox{-.1em}{\tiny\it FS}}}}}

\newcommand{\forsimpreo}{\sqsubseteq_{S}}
\newcommand{\forsimequiv}{\simeq_{S}}
\newcommand{\failsimpreo}{\sqsubseteq_{\it FS}}

\newcommand{\failsimequiv}{\simeq_{\it FS}}

\newcommand{\modalFSpreo}{\mathrel{\sqsubseteq}^{\FSL}\!}
\newcommand{\modalSpreo}{\mathrel{\sqsubseteq}^{\SL}\!}
\newcommand{\tdc}{\sqsubseteq_{\it TD}}
\newcommand{\fdc}{\sqsubseteq_{\it FD}}

\newcommand{\dist}[1]{\mathop{\mbox{$\mathcal D$}} ({#1})   } % distributions
\newcommand{\pdist}[1]{\overline{#1}  } % distributions
\newcommand{\support}[1]{\lceil{#1}\rceil}
\newcommand{\lift}[1]{\mathrel{\overline{#1}}}

\newcommand{\FSL}{\mathcal{F}}
\newcommand{\SL}{\mathcal{L}}
\newcommand{\diam}[1]{\langle#1\rangle}
\newcommand{\refuse}[1]{\mathop{\mathbf{ref}(#1)}}

\newcommand{\setof}[2]{\{ \, #1 \, \mid \, #2 \, \}}% set comprehension
\newcommand{\sset}[1]{\{ {#1}  \}  } % singleton set

\DeclareSymbolFont{smallcaps}{\encodingdefault}{\rmdefault}{m}{sc}
\DeclareSymbolFontAlphabet\mathsc{smallcaps}
\newcommand{\Rulef}[1]{{\mathsc{#1}}}   %%font for rules
\newcommand{\Rlts}[1]{\ensuremath{\Rulef{(#1)}}}%%%%

\newcommand{\Ename}[1]{\ensuremath{\mathbf{(#1)}}}%%%%

\makeatletter
\newbox\@topbox
\newbox\@botbox
\newcommand{\linfer}[3][{}]{%
    \setbox\@topbox\hbox{%
      \renewcommand{\arraystretch}{1}%
      $\begin{array}{l}#2\end{array}$}%
  \setbox\@botbox\hbox{%
    \def\tilde##1{\widetilde{##1}}%
    \renewcommand{\arraystretch}{1}%
    $\begin{array}{l}#3\end{array}$}%
  \ifdim\wd\@botbox<\wd\@topbox% which is bigger?  set the other to match
    \setbox\@botbox\hbox to \wd\@topbox{\box\@botbox\hfil}
  \else
    \setbox\@topbox\hbox to \wd\@botbox{\box\@topbox\hfil}
  \fi
  \frac{\box\@topbox}{\box\@botbox}
  }
\newcommand{\linferSIDE}[4][{}]{%
    \setbox\@topbox\hbox{%
      \renewcommand{\arraystretch}{1}%
      $\begin{array}{lr@{}}#2&#4\end{array}$}%
  \setbox\@botbox\hbox{%
    \def\tilde##1{\widetilde{##1}}%
    \renewcommand{\arraystretch}{1}%
    $\begin{array}{l}#3\end{array}$}%
  \ifdim\wd\@botbox<\wd\@topbox% which is bigger?  set the other to match
    \setbox\@botbox\hbox to \wd\@topbox{\box\@botbox\hfil}
  \else
    \setbox\@topbox\hbox to \wd\@botbox{\box\@topbox\hfil}
  \fi
  \frac{\box\@topbox}{\box\@botbox}
  }
\newcommand{\slinfer}[2][{}]{\renewcommand{\arraystretch}{1}%
                  \begin{array}{l}#2\end{array}}

\def\mine{\@transition\mapstofill}
\def\goesto{\@transition\rightarrowfill}
\def\Goesto{\@transition\Rightarrowfill}
\def\ngoesto{\@transition\lnrightarrowfill} % changed by Rob 080112
\def\nsgoesto{\@transition\nmapstofill}
\def\comesfrom{\@transition\leftarrowfill}
\def\nGoesto{\@transition\nRightarrowfill}
\def\@transition#1{\@ifnextchar[{\@@transition{#1}}{\@@transition{#1}[]}}
\newbox\@transbox
\newbox\@arrowbox
\def\mapstofill{$\m@th\mathord\mapstochar\mathord-\mkern-6mu%
  \cleaders\hbox{$\mkern-2mu\mathord-\mkern-2mu$}\hfill
  \mkern-6mu\mathord\rightarrow$}
\def\Rightarrowfill{$\m@th\mathord=\mkern-6mu%
  \cleaders\hbox{$\mkern-2mu\mathord=\mkern-2mu$}\hfill
  \mkern-6mu\mathord\Rightarrow$}
\def\lnrightarrowfill{$\m@th\mathord-\mkern-6mu%
  \cleaders\hbox{$\mkern-2mu\mathord-\mkern-2mu$}\hfill
  \mkern-6mu\mathord\not\mkern-2mu\mathord\rightarrow$}
\def\nrightarrowfill{$\m@th\mathord-\mkern-6mu%
  \cleaders\hbox{$\mkern-2mu\mathord-\mkern-2mu$}\hfill
  \mkern-6mu\mathord\not\mathord-\mkern-6mu
  \cleaders\hbox{$\mkern-2mu\mathord-\mkern-2mu$}\hfill
  \mkern-6mu\mathord\rightarrow$}
\def\nmapstofill{$\m@th\mathord\mapstochar\mathord-\mkern-6mu%
  \cleaders\hbox{$\mkern-2mu\mathord-\mkern-2mu$}\hfill
  \mkern-6mu\mathord\not\mathord-\mkern-6mu
  \cleaders\hbox{$\mkern-2mu\mathord-\mkern-2mu$}\hfill
  \mkern-6mu\mathord\rightarrow$}
\def\nRightarrowfill{$\m@th\mathord=\mkern-6mu%
  \cleaders\hbox{$\mkern-2mu\mathord=\mkern-2mu$}\hfill
  \mkern-6mu\mathord\not\mathord=\mkern-6mu
  \cleaders\hbox{$\mkern-2mu\mathord=\mkern-2mu$}\hfill
  \mkern-6mu\mathord\Rightarrow$}
\def\@@transition#1[#2]%
   {\setbox\@transbox\hbox
       {\vrule height 1.5ex depth .7ex width 0ex\hskip0.25em$\scriptstyle#2$\hskip0.25em}
   \ifdim\wd\@transbox<1.5em
      \setbox\@transbox\hbox to 1.5em{\hfil\box\@transbox\hfil}\fi
   \setbox\@arrowbox\hbox to \wd\@transbox{#1}
   \ht\@arrowbox\z@\dp\@arrowbox\z@
   \setbox\@transbox\hbox{$\mathop{\box\@arrowbox}\limits^{\box\@transbox}$}
   \ht\@transbox\z@\dp\@transbox\z@
   \mathrel{\box\@transbox}}
\def\nxmapsto{\@transition\nmapstofill}
\def\xmapsto{\@transition\mapstofill}
\def\mapstofill{$\m@th\mathord\mapstochar\mathord-\mkern-6mu%
  \cleaders\hbox{$\mkern-2mu\mathord-\mkern-2mu$}\hfill
  \mkern-6mu\mathord\rightarrow$}
\def\nmapstofill{$\m@th\mathord\mapstochar\mathord-\mkern-6mu%
  \cleaders\hbox{$\mkern-2mu\mathord-\mkern-2mu$}\hfill
  \mkern-6mu\mathord\arrownot\mathord-\mkern-6mu
  \cleaders\hbox{$\mkern-2mu\mathord-\mkern-2mu$}\hfill
  \mkern-6mu\mathord\rightarrow$}
\makeatother

\newcommand{\Ar}[1]{\mathbin{\stackrel{#1}{\longrightarrow}}}
\newcommand{\ar}[1]{\mathrel{\goesto[{#1}]}}
\newcommand{\nar}[1]{\mathrel{\ngoesto[{#1\;}]}}
\newcommand{\dar}[1]{\mathrel{\Goesto[\raisebox{.08em}{\scriptsize$#1$}]}} %weak arrow
\newcommand{\hatar}[1]{\mathrel{\goesto[\hat{{#1}}]}}

\newcommand{\PrR}{{\rm Pr_R}}
\newcommand{\WW}{{\mathbb W}}
\newcommand{\eR}{{\rm R}}           % a fully probabilistic automaton
\newcommand{\depth}{\mathop{\mathit{depth}}}

\newcommand\Def[1] {Definition~\ref{#1}}
\newcommand\Thm[1] {Theorem~\ref{#1}}
\newcommand\Eqn[1] {(\ref{#1})}
\newcommand\Lem[1] {Lemma~\ref{#1}}
\newcommand\Fig[1] {Figure~\ref{#1}}
\newcommand\Prop[1] {Proposition~\ref{#1}}
\newcommand\Sect[1] {Section~\ref{#1}}

\newcommand{\MustAxleq}{\mathrel{\sqsubseteq_{E_\textrm{\rm must}}}}
\newcommand{\Eeq}{\mathrel{=_{E}}}
\newcommand{\Peq}{\mathrel{=_{\textrm{\rm prob}}}}
\newcommand{\MayAxleq}{\mathrel{\sqsubseteq_{E_\textrm{\rm may}}}}
\newcommand{\nCSP}{\ensuremath{\mathsf{nCSP}}\xspace}
\newcommand{\inits}[1]{\mathrel{\textit{inits}(#1)}}
\renewcommand{\max}{\mathop{\mathit{max}}}
\renewcommand{\min}{\mathop{\mathit{min}}}

\def\doi{4 (4:4) 2008}
\lmcsheading%
{\doi}
{1--33}
{}
{}
{Feb.~\phantom{0}7, 2008}
{Oct.~28, 2008}
{}   

\begin{document}

\title[Characterising Testing Preorders for Finite Probabilistic Processes]
      {Characterising Testing Preorders\\ for Finite Probabilistic
      Processes\rsuper *}

\author[Y.~Deng]{Yuxin Deng\rsuper a}
\address{{\lsuper a}Shanghai Jiao Tong University, China, and University of New South Wales, Australia}
\email{yuxindeng@sjtu.edu.cn}
\thanks{{\lsuper{a,c,d}}Deng was supported by the National Natural Science
  Foundation of China (60703033) and the ARC (DP034557). 
  Hennessy would like to acknowledge the support of SFI and The Royal
Society, UK.  Morgan would like to
  acknowledge the support of the Australian Research Council (ARC) Grant DP034557.}

\author[R.~van Glabbeek]{Rob van Glabbeek\rsuper b}
\address{{\lsuper b}National ICT Australia, and University of New South Wales, Australia}
\email{rvg@cs.stanford.edu}

\author[M.~Hennessy]{Matthew Hennessy\rsuper c}
\address{{\lsuper c}Trinity College Dublin, Ireland}
\email{matthew.hennessy@cs.tcd.ie}

\author[C.~Morgan]{Carroll Morgan\rsuper d}
\address{{\lsuper d}University of New South Wales, Australia}
\email{carrollm@cse.unsw.edu.au}

\titlecomment{{\lsuper *}An extended abstract of this paper has appeared as \cite{DGHMZ07a}.}

\begin{abstract}
In 1992 Wang \& Larsen extended the may- and must preorders of De
Nicola and Hennessy to processes featuring probabilistic as well as
nondeterministic choice. They concluded with two problems that have
remained open throughout the years, namely to find complete
axiomatisations and alternative characterisations for these preorders.
This paper solves both problems for finite processes with silent
moves. It characterises the may preorder in terms of simulation, and
the must preorder in terms of failure simulation. It also gives a
characterisation of both preorders using a modal logic. Finally it
axiomatises both preorders over a probabilistic version of finite CSP.
\end{abstract}

\maketitle
%-------------------------------------------------------------------------
\section{Introduction}

\noindent
A satisfactory semantic theory for processes which encompass both
nondeterministic and probabilistic behaviour has been a long-standing
research problem
\cite{HJ90,WL92,Low93,JHW94,SL94,Sei95,Seg95,JW95,MMSS96,Seg96,HSM97,KN98,Mis00,BS01,JW02,LSV03,CCVP03,TKP05,DP07}.
In 1992 Wang \& Larsen posed the problems of finding complete axiomatisations
and alternative characterisations for a natural generalisation of the
standard testing preorders \cite{DNH84} to such processes \cite{WL92}.
Here we solve both problems, at least for finite processes, by providing
a detailed account of both may- and must testing preorders for a
finite version of the process calculus CSP extended with
probabilistic choice. For each preorder we provide three independent
characterisations, using\linebreak[1]
(i) co-inductive simulation relations, \,
(ii) a modal logic \, and
(iii) sets of inequations.

\paragraph*{\bf Testing processes:}
Our starting point is the finite process calculus \pCSP \cite{DGHMZ07}
obtained by adding a probabilistic choice operator to finite CSP;
like others who have done the same, we now have \emph{three} choice
operators, external $P \extc Q$, internal  $P \mathbin{\intc} Q$ and the newly added
probabilistic choice $P \mathbin{\probc{p}} Q$. So a semantic theory for \pCSP
will have to provide a coherent account of the precise relationships
between these operators.

As a first step, in \Sect{sec:pcsp} we provide an interpretation of
\pCSP as a \emph{probabilistic labelled transition system}, in which,
following \cite{SL94,JHW94}, state-to-state transitions like $s \Ar{\alpha} s'$ from
standard labelled transition systems are generalised to the form $s
\ar{\alpha} \Delta$, where $\Delta$ is a \emph{distribution}, a mapping
assigning probabilities to states. With this interpretation we
obtain in \Sect{sec:testing} a version of the testing preorders of \cite{DNH84} for \pCSP
processes, $\pMayleq$ and $\pMustleq$.
These are based on the ability of processes to pass \emph{tests}; the
tests we use are simply \pCSP processes in which certain \emph{states}
are marked as \emph{success states}. See \cite{DGHMZ07} for a
detailed discussion of the power of such tests.

The object of this paper is to give alternative characterisations of these
testing preorders. This problem was addressed previously by Segala
in \cite{Seg96}, but using testing preorders ($\pdmMayleq$ and
$\pdmMustleq$) that differ in two ways from the ones in
\cite{DNH84,henn,WL92,DGHMZ07} and the present paper. First of all, in
\cite{Seg96} the success of a test is achieved by the \emph{actual
execution} of a predefined \emph{success action}, rather than the
reaching of a success state.  We call this an \emph{action}-based approach, as
opposed to the \emph{state}-based approach used in this paper.
Secondly, \cite{Seg96} employs a countable number of success actions instead of a single one;
we call this \emph{vector-based}, as opposed to \emph{scalar},
testing. Segala's results in \cite{Seg96} depend crucially on this
form of testing. To achieve our current results, we need Segala's
preorders as a stepping stone. We relate them to ours by considering
intermediate preorders $\pdMayleq$ and \plat{$\pdMustleq$} that arise from
action-based but scalar testing, and use a recent result \cite{DGMZ07}
saying that for finite processes the preorders $\pdmMayleq$ and
$\pdmMustleq$ coincide with \plat{$\pdMayleq$} and \plat{$\pdMustleq$}.
Here we show that on {\pCSP} the preorders \plat{$\pdMayleq$} and
\plat{$\pdMustleq$} also coincide with $\pMayleq$ and $\pMustleq$.%
\footnote{However in the presence of divergence they are slightly different.}

\paragraph*{\bf Simulation preorders:} In \Sect{sec:simulations} we
use the transitions $s \Ar{\alpha} \Delta$ to define two co-inductive
preorders, the \emph{simulation} preorder $\forsimpreo$
\cite{Seg95,LSV03,DGHMZ07}, and the novel
\emph{failure simulation} preorder $\failsimpreo$ over \pCSP processes.
The latter extends the failure simulation preorder of \cite{vG93} to probabilistic processes.
Their definition uses a natural generalisation of the
transitions, first (Kleisli-style) to take the form $\Delta \Ar{\alpha} \Delta'$, and then to
\emph{weak} versions $\Delta \dar{\alpha} \Delta'$. The second
preorder differs from the first one in the use of a \emph{failure}
predicate $s \nar{X}$, indicating that in the state $s$ none of the
actions in $X$ can be performed.

Both preorders are preserved by all the operators
in \pCSP, and are \emph{sound} with respect to the testing
preorders; that is $P \forsimpreo Q$ implies $P \pMayleq Q$
and $P \failsimpreo Q$ implies $P \pMustleq Q$.
For $\forsimpreo$ this was established in \cite{DGHMZ07}, and here
we use similar techniques in the proofs for $\failsimpreo$.
But \emph{completeness}, that the
testing preorders imply the respective simulation preorders, requires
some ingenuity. We prove it indirectly, involving a characterisation of
the testing and simulation preorders in terms of a modal logic.

\paragraph*{\bf Modal logic:} Our modal logic, defined in
\Sect{sec:modal-logic}, uses finite conjunction $\bigwedge_{i\in I}\phi_i$,
the modality $\diam{a}\phi$ from the Hennessy-Milner Logic \cite{HM85},
and a novel probabilistic construct $\bigoplus_{i\in I}p_i \cdot \phi_i$.
A satisfaction relation between processes and formulae then gives, in a
natural manner, a \emph{logical preorder} between processes:
$P \modalSpreo Q$ means that every $\SL$-formula satisfied by $P$ is
also satisfied by $Q$. We establish that $\modalSpreo$
coincides with $\forsimpreo$ and $\pMayleq$.

To capture failures, we add, for every set of actions $X$, a formula
$\refuse{X}$ to our logic, satisfied by any process which, after it
can do no further internal actions, can perform none of the actions in
$X$ either. The constructs $\bigwedge$, $\diam{a}$ and $\refuse{}$
stem from the modal characterisation of the non-probabilistic failure
simulation preorder, given in \cite{vG93}. We show that $\pMustleq$,
as well as $\failsimpreo$, can be characterised in a similar manner
with this extended modal logic.

\paragraph*{\bf Proof strategy:}
We prove these characterisation results through two cycles of inclusions:
$$\begin{array}{@{}l@{\;\,\;}c@{\;\,\;}l@{\;\,\;}c@{\;\,\;}l@{\;\,\;}c@{\;\,\;}l@{\;\,\;}c@{\;\,\;}l@{\;\,\;}c@{\;\,\;}l@{}}
  \modalSpreo  &\subseteq&  \forsimpreo  &\stackrel{\mbox{\scriptsize\cite{DGHMZ07}}}{\subseteq}&  \pMayleq  &\subseteq&  \pdMayleq  &\stackrel{\mbox{\scriptsize\cite{DGMZ07}}}{=}&  \pdmMayleq  &\subseteq&  \modalSpreo\\[1pt]
  \modalFSpreo &\subseteq&  \failsimpreo &\subseteq&  \pMustleq &\subseteq&  \pdMustleq &\stackrel{\mbox{\scriptsize\cite{DGMZ07}}}{=}&  \pdmMustleq &\subseteq&  \modalFSpreo\\[-6pt]
\multicolumn{2}{@{}c@{\;\,\;}}{\underbrace{\hspace{30pt}}}&
\multicolumn{2}{@{}c@{\;\,\;}}{\underbrace{\hspace{35pt}}}&
\underbrace{\hspace{32pt}}&
\multicolumn{2}{@{}c@{\;\,\;}}{\underbrace{\hspace{47pt}}}&
\multicolumn{2}{@{}c@{\;\,\;}}{\underbrace{\hspace{51pt}}}&
\multicolumn{2}{@{}c@{}}{\underbrace{\hspace{31pt}}}\\[-1pt]
\multicolumn{2}{@{}c@{\;\,\;}}{\mbox{\small\sl Sec.~\ref{sec:modal-logic}}} &
\multicolumn{2}{@{}c@{\;\,\;}}{\mbox{\small\sl Sec.~\ref{sec:simulations}}} &
  \mbox{\small\sl Sec.~\ref{sec:testing}} &
\multicolumn{2}{@{}c@{\;\,\;}}{\mbox{\small\sl Sec.~\ref{sec:state-vs-action}}} &
\multicolumn{2}{@{}c@{\;\,\;}}{\mbox{\small\sl Sec.~\ref{sec:vector}}} &
\multicolumn{2}{@{}c@{}}{\mbox{\small\sl Sec.~\ref{sec:chartests}}}
\end{array}$$
In \Sect{sec:modal-logic} we show that $P \modalSpreo Q$
implies $P \forsimpreo Q$ (and hence $P \pMayleq Q$), and likewise for
$\modalFSpreo$ and $\failsimpreo$; the proof involves constructing,
for each \pCSP process $P$, a \emph{characteristic formula} $\phi_P$.
To obtain the other direction, in \Sect{sec:chartests} we show how
every modal formula $\phi$ can be captured, in some sense, by a test
$T_\phi$; essentially the ability of a \pCSP process to satisfy $\phi$
is determined by its ability to pass the test $T_\phi$. We capture
the conjunction of two formulae by a probabilistic choice between the
corresponding tests; in order to prevent the results from these tests
getting mixed up, we employ the vector-based tests of \cite{Seg96}, so
that we can use different success actions in the separate probabilistic branches.
Therefore, we complete our proof by demonstrating that the state-based testing
preorders imply the action-based ones (\Sect{sec:state-vs-action})
and recalling the result from \cite{DGMZ07} that the action-based scalar
testing preorders imply the vector-based ones (\Sect{sec:vector}).

\paragraph*{\bf (In)equations:}
It is well-known that may- and must testing for standard CSP can be
captured equationally \cite{DNH84,BHR84,henn}. In \cite{DGHMZ07} we
showed that most of the standard equations are no longer valid in
the probabilistic setting of \pCSP; we also provided
a set of axioms which are complete with respect to
(probabilistic) may-testing for the sub-language of \pCSP without
probabilistic choice. Here we extend this result, by
showing, in \Sect{sec:inequations}, that both $P \pMayleq Q$ and $P
\pMustleq Q$ can still be captured equationally over full \pCSP.
In the may case the essential (in)equation required is
\begin{equation*}
 a.(P \probc{p} Q) \Wide{\sqsubseteq} a.P \probc{p} a.Q
\end{equation*}
The must case is more involved: in the absence of the
distributivity of the external and internal choices over each other,
to obtain completeness we require a complicated inequational schema.

%-------------------------------------------------------------------------
\section{Finite probabilistic CSP}\label{sec:pcsp}

\noindent
%% The set of actions is required to be finite, because later on we
%% make a characteristic test (a finite PCSP expression) out of ref(X)
%% by means of an external choice of all actions that cannot be done.
Let $\Act$ be a finite set of \emph{visible} (or \emph{external})
actions, ranged over by $a,b,\cdots$, which
processes can perform.  Then the finite probabilistic CSP
processes are given by the following two-sorted syntax:
\begin{eqnarray*}
  P &::=&  S \BNFsep P \probc{p} P
\\
  S &::=&  \Cnil \BNFsep a.P \BNFsep P \intc P \BNFsep S \extc S \BNFsep S \Par{A} S
\end{eqnarray*}
We write \pCSP, ranged over by $P,Q$, for the set of process terms
defined by this grammar, and \sCSP, ranged over by $s,t$, for
the subset comprising only the state-based process terms (the sub-sort
$S$ above).

The process $P \probc{p} Q$, for $0< p < 1$,
represents a \emph{probabilistic choice} between $P$ and $Q$: with
probability $p$ it will act like $P$ and with probability $1\Minus p$ it
will act like $Q$.  Any process is a probabilistic combination of
state-based processes
built by repeated application of the operator $\probc{p}$. The
state-based processes have a CSP-like syntax, involving the
stopped process $\Cnil$, action prefixing $a.\_\!\_$ for $a\in\Act$,
\emph{internal-} and \emph{external choices} $\intc$ and $\extc$,
and a \emph{parallel composition} $\Par{A}$ for $A\subseteq \Act$.

The process $P\intc Q$ will first do a so-called \emph{internal action} $\tau\mathop{\not\in}\Act$,
choosing \emph{nondeterministically} between $P$ and $Q$.
Therefore $\intc$, like $a.\_\!\_$\,, acts as a \emph{guard}, in the
sense that it converts any process arguments into a state-based process.

The process $s\extc t$ on the other hand does not perform actions itself,
but merely allows its arguments to proceed, disabling one argument as
soon as the other has done a visible action. In order for this process
to start from a state rather than a probability distribution of states,
we require its arguments to be state-based as well; the
same applies to $\Par{A}$.

Finally, the expression $s \Par{A} t$, where $A \subseteq \Act$,
represents processes $s$ and $t$ running in parallel. They may
synchronise by performing the same action from $A$ simultaneously; such a
synchronisation results in $\tau$.
In addition $s$ and $t$ may independently do any action from
$(\Act\mathord{\setminus} A)\cup\{\tau\}$.

Although formally the operators $\extc$ and $\Par{A}$ can only be applied to state-based
processes, informally we use expressions of the form  $P\extc Q$ and $P\Par{A} Q$, where  $P$
and $Q$ are \emph{not} state-based, as syntactic
sugar for  expressions in the above syntax obtained by
distributing $\extc$ and $\Par{A}$ over $\probc{p}$.
Thus for example $s \extc (t_1 \probc{p} t_2)$ abbreviates the term
$(s \extc t_1) \probc{p} (s \extc t_2)$.

The full language of CSP \cite{BHR84,csp,OH86} has many more
operators; we have simply chosen a representative selection, and have added
probabilistic choice.
Our parallel operator is not a CSP primitive, but it can easily
be expressed in terms of them---in particular $P\Par{A}Q
= (P\|_AQ)\backslash A$, where $\|_A$ and $\backslash A$ are the
parallel composition and hiding operators of \cite{OH86}.  It can also
be expressed in terms of the parallel composition, renaming and
restriction operators of CCS\@.  We have chosen this (non-associative)
operator for convenience in defining the application of tests to processes.

As usual we may elide $\Cnil$; the prefixing
operator $a.\_\!\_$ binds stronger than any binary operator; and
precedence between binary operators is indicated via brackets
or spacing. We will also sometimes use indexed binary
operators, such as $\bigoplus_{i\in I}p_i\Cdot P_i$ with
$\sum_{i\in I}p_i=1$ and all $p_i>0$, and $\bigbox_{i \in I}P_i$,
for some finite index set $I$.

The above intuitions are formalised by an \emph{operational semantics}%
\footnote{Although the syntax of \pCSP is similar to other
probabilistic extensions of CSP \cite{Low93,MMSS96,Mis00}, our
semantics differs.  For more detailed comparisons, see
Section~\ref{sec:conc}.}  associating with each process term a
graph-like structure representing its possible reactions to users'
requests: we use a generalisation of labelled transition systems
\cite{ccs} that includes probabilities.

A (discrete) probability distribution over a set $S$ is a function
{$\Delta\!: S \rightarrow [0,1]$} with \mbox{$\sum_{s\in S}\! \Delta(s)=1$};
the \emph{support} of $\Delta$ is given by
$\support{\Delta} = \setof{s \inp S}{\Delta(s) > 0}$.
%% We only deal with distributions with finite support, because for
%% other ones we can't define a characteristic formula whose
%% characteristic test is a finite pCSP expression.
We write $\dist{S}$, ranged over by $\Delta, \Theta,\Phi$, for the set
of all  distributions over $S$ with finite support; these finite distributions
are sufficient for the results of this paper. We also write
$\pdist{s}$ to denote the point distribution assigning probability 1
to $s$ and 0 to all others, so that $\support{\pdist{s}} = \{s\}$.
If $p_i\geq 0$ and $\Delta_i$ is a distribution for each
$i$ in some finite index set $I$, and $\sum_{i\in I}p_i = 1$, then the probability distribution
$\sum_{i \in I}p_i \cdot \Delta_i \in \dist{S}$ is given by
\begin{equation*}
  (\sum_{i \in I}p_i \cdot \Delta_i)(s) ~~~=~~~ \sum_{i \in I} p_i \cdot\Delta_i(s) \;;
\end{equation*}
we will sometimes write it as $p_1 \cdot \Delta_1 + \ldots + p_n \cdot \Delta_n$
when the index set $I$ is $\sset{1,\ldots,n}$.

For $\Delta$ a distribution over $S$ and function
\mbox{$f\!:S\mathbin\rightarrow X$} into a vector space $X$ we
sometimes write $\Exp{\Delta}(f)$ for $\sum_{s\in S} \Delta(s) \cdot f(s)$,
the \emph{expected value} of $f$.  Our primary use of this notation is
with $X$ being the vector space of reals or tuples of reals.
More generally, for function $F: S \rightarrow \subs{X}$ with
$\subs{X}$ being the collection of non-empty subsets of $X$, we define
$\Exp{\Delta}{F} \Defs \setof{\Exp{\Delta}(f)}{f \IN F}$; here $f  \IN F$ means that
$f\!:S\mathbin\rightarrow X$ is a \emph{choice function} for $F$, that is it
satisfies the constraint that  $f(s)\inp F(s)$ for all $s\inp S$.

We now give the probabilistic generalisation of labelled transition systems (LTSs):

\begin{defi}
A \emph{probabilistic labelled transition system}
(pLTS)\footnote{Essentially the same model has appeared in the
literature under different names such as \emph{NP-systems}
\cite{JHW94}, \emph{probabilistic processes} \cite{JW95},
\emph{simple probabilistic automata} \cite{Seg95},
\emph{probabilistic transition systems} \cite{JW02} etc.
Furthermore, there are strong structural similarities with \emph{Markov
  Decision Processes} \cite{Put94,DGMZ07}.}
 is a triple
$\langle S, L, \rightarrow \rangle$, where
\begin{enumerate}[(i)]

\item $S$ is a set of states,

\item $L$ is a set of transition labels,

\item relation $\rightarrow$ is a subset of $S \times L \times \dist{S}$.

\end{enumerate}
As with LTSs, we usually write $s \Ar{\alpha} \Delta$ for
$(s,\alpha,\Delta) \inp \mathord\rightarrow$,
$s\ar{\alpha}$ for $\exists\Delta: s\Ar{\alpha}\Delta$ and
$s\!\rightarrow$ for $\exists\alpha\!: s\,\Ar{\alpha}$.
An LTS may be viewed as a degenerate pLTS, one in which only point
distributions are used.
\end{defi}

\begin{figure}[t]
\vspace{-1em}
  \begin{align*}
    &\slinfer[\Rlts{action}]{a.P \ar{a} \Op{P}}
\\[3pt]
    & \slinfer[\Rlts{int.l}]{P \intc Q \ar{\tau} \Op{P}}
    && \slinfer[\Rlts{int.r}]{P \intc Q \ar{\tau} \Op{Q}}
\\
  &\linfer[\Rlts{ext.l}]{s_1 \ar{a} \Delta}
                {s_1 \extc s_2 \ar{a} \Delta}
  &&
 \linfer[\Rlts{ext.r}]{s_2 \ar{a} \Delta}
                {s_1 \extc s_2 \ar{a} \Delta}
\\
   &\linfer[\Rlts{ext.i.l}]{s_1 \ar{\tau} \Delta}
                 {s_1 \extc s_2 \ar{\tau} \Delta \extc {s_2}}
   && \linfer[\Rlts{ext.i.r}]{s_2 \ar{\tau} \Delta}
                 {s_1 \extc s_2 \ar{\tau} {s_1} \extc \Delta}
\\
  &\linferSIDE[\Rlts{par.l}]{s_1 \ar{\alpha} \Delta}
                 {s_1 \Par{A} s_2 \ar{\alpha} \Delta \Par{A} {s_2}}
                 {~~~~~\alpha \mathop{\not\in} A}
   && \linferSIDE[\Rlts{par.r}]{s_2 \ar{\alpha} \Delta}
                 {s_1 \Par{A} s_2 \ar{\alpha} {s_1} \Par{A} \Delta}
                 {~~~~~\alpha \mathop{\not\in} A}
\\
  &\makebox[4.5cm][l]{$\linferSIDE[\Rlts{par.i}]{s_1 \ar{a} \Delta_1,\;s_2 \ar{a} \Delta_2}
                {s_1 \Par{A} s_2 \ar{\tau} \Delta_1 \Par{A} \Delta_2}
        {a\inp A}$}
 \end{align*}
\caption{Operational semantics of $\mathsf{pCSP}$\label{fig:opsem}}
\end{figure}

\noindent
The operational semantics of \pCSP is defined by
a particular pLTS  $\langle \sCSP, \Act_\tau, \rightarrow \rangle$,
constructed by taking \sCSP to be the set of states
and $\Act_\tau := \Act \cup \{\tau\}$ the set of transition labels;
we let $a$ range over $\Act$ and $\alpha$ over $\Act_\tau$.
We interpret \pCSP processes $P$ as distributions $\Op{P} \in \dist{\sCSP}$
via the function $\Op{\_\!\_}:\pCSP \rightarrow \dist{\sCSP}$ defined below:
$$\begin{array}{r@{~~\Defs~~}l}
\Op{s} & \pdist{s} ~~~~~~\mbox{for } s \inp\sCSP\\
\Op{P \probc{p} Q} & p\cdot\Op{P} + (1-p)\cdot \Op{Q}\;.
\end{array}$$
Note that for each $P\in\pCSP$ the distribution $\Op{P}$ is finite,
that is it has finite support.
The definition of the relations $\ar{\alpha}$ is given in
\Fig{fig:opsem}.
These rules are very similar to the
standard ones used to interpret CSP as an LTS \cite{OH86}, but modified
so that the result of an action is a distribution.  The rules for
external choice and parallel composition use an obvious notation for distributing
an operator over a distribution; for example $\Delta \extc s$ represents the
distribution given by
\begin{displaymath}
  (\Delta \extc s)(t) =
  \begin{cases}
    \Delta(s')  &\text{if $t = s' \extc s$}\\
     0           &\text{otherwise}.
  \end{cases}
\end{displaymath}
We sometimes write $\tau.P$ for $P\intc P$, thus giving $\tau.P \ar{\tau} \Op{P}$.

We graphically depict the operational semantics of a \pCSP expression
$P$ by drawing the part of the pLTS defined above that is reachable
from $\Op{P}$ as a finite acyclic directed graph, often unwound into a tree.
States are represented by nodes of the
form $\bullet$ and distributions by nodes of the form $\circ$. For any
state $s$ and distribution $\Delta$ with $s\ar{\alpha}\Delta$ we draw
an edge from $s$ to $\Delta$, labelled with $\alpha$.
For any distribution $\Delta$ and state
$s$ in $\support{\Delta}$, the support of $\Delta$, we draw an edge from
$\Delta$ to $s$, labelled with $\Delta(s)$.

\begin{exa}\label{ex:1}
Consider the two processes
\[\begin{array}{rcl}
P & := & a.((b.d\ \Box\ c.e)\probc{\frac{1}{2}}(b.f\ \Box\ c.g)) \\
Q & := & a.((b.d\ \Box\ c.g)\probc{\frac{1}{2}}(b.f\ \Box\ c.e)).
\end{array}\]
Their tree representations are depicted in Figure~\ref{fig:examples} (i) and (ii).
To make these trees more compact we omit  nodes
$\circ$ when they represent trivial point distributions.
\end{exa}

\begin{figure}
\vspace{-.7em}
\[\begin{array}{ccc}
    \psset{arrows=->}
    \pstree[nodesep=0pt,levelsep=20pt,labelsep=.05]{\Tdot}
     {\pstree{\Tdot[dotstyle=o] \tlput{a}}
             {\pstree{\Tdot
                 \tlput[tpos=.1]{\scriptscriptstyle{\frac{1}{2}}}}
                 {\pstree{\Tdot \tlput{b}}
                    {\pstree{\Tdot \tlput{d}}{}}
                  \pstree{\Tdot \trput{c}}
                    {\pstree{\Tdot \trput{e}}{}}
                 }
              \pstree{\Tdot
                 \trput[tpos=.1]{\scriptscriptstyle{\frac{1}{2}}}}
                 {\pstree{\Tdot \tlput{b}}
                    {\pstree{\Tdot \tlput{f}}{}}
                  \pstree{\Tdot \trput{c}}
                    {\pstree{\Tdot \trput{g}}{}}
                 }
             }
     }

& \qquad\qquad\qquad
    \psset{arrows=->}
    \pstree[nodesep=0pt,levelsep=20pt,labelsep=.05]{\Tdot}
     {\pstree{\Tdot[dotstyle=o] \tlput{a}}
             {\pstree{\Tdot
                 \tlput[tpos=.1]{\scriptscriptstyle{\frac{1}{2}}}}
                 {\pstree{\Tdot \tlput{b}}
                    {\pstree{\Tdot \tlput{d}}{}}
                  \pstree{\Tdot \trput{c}}
                    {\pstree{\Tdot \trput{g}}{}}
                 }
              \pstree{\Tdot
                 \trput[tpos=.1]{\scriptscriptstyle{\frac{1}{2}}}}
                 {\pstree{\Tdot \tlput{b}}
                    {\pstree{\Tdot \tlput{f}}{}}
                  \pstree{\Tdot \trput{c}}
                    {\pstree{\Tdot \trput{e}}{}}
                 }
             }
     }

& \qquad\qquad\qquad
    \psset{arrows=->}
    \pstree[nodesep=0pt,levelsep=20pt,labelsep=.05]{\Tdot}
     {\pstree{\Tdot \trput{a}}
             {\pstree{\Tdot[dotstyle=o] \tlput{\tau}}
                {\pstree{\Tdot \tlput[tpos=.1]{\scriptscriptstyle{\frac{1}{2}}}}
                 {\pstree{\Tdot \tlput{b}}
                    {\pstree{\Tdot \tlput{d}}
                     {\pstree{\Tdot \tlput{\omega}}{}}
                    }
                 }
                 \pstree{\Tdot \trput[tpos=.1]{\scriptscriptstyle{\frac{1}{2}}}}
                  {\pstree{\Tdot \trput{c}}
                    {\pstree{\Tdot \trput{e}}
                      {\pstree{\Tdot \trput{\omega}}{}}
                    }
                  }
                 }
               \pstree{\Tdot[dotstyle=o] \trput{\tau}}
                {\pstree{\Tdot \tlput[tpos=.1]{\scriptscriptstyle{\frac{1}{2}}}}
                 {\pstree{\Tdot \tlput{b}}
                    {\pstree{\Tdot \tlput{f}}
                     {\pstree{\Tdot \tlput{\omega}}{}}
                    }
                 }
                 \pstree{\Tdot \trput[tpos=.1]{\scriptscriptstyle{\frac{1}{2}}}}
                  {\pstree{\Tdot \trput{c}}
                    {\pstree{\Tdot \trput{g}}
                      {\pstree{\Tdot \trput{\omega}}{}}
                    }
                  }
                 }
             }
     }
\\
(i)\;\; P %a.((b.d\ \Box\ c.e)\probc{\frac{1}{2}}(b.f\ \Box\ c.g))
& \qquad\qquad\qquad (ii)\;\; Q %a.((b.d\ \Box\ c.g)\probc{\frac{1}{2}}(b.f\ \Box\ c.e))
& \qquad\qquad\qquad (iii)\;\; T %a.((b.d.\eureka \probc{\frac{1}{2}} c.e.\eureka) \intc (b.f.\eureka\probc{\frac{1}{2}} c.g.\eureka))
\end{array}\]
\caption{Example processes $P,Q$ and test $T$}\label{fig:examples}
\end{figure}

%-------------------------------------------------------------------------
\section{Testing \pCSP processes}\label{sec:testing}

\noindent
A \emph{test} is a \pCSP process except that it may have subterms
$\eureka.P$ for fresh $\eureka\mathop{\not\in}\Act_\tau$, a special action
reporting success; we write \pCSPw for the set of all tests, and
\sCSPw for the subset of state-based process terms that may involve
the action $\eureka$, and the operational semantics above
is extended by treating $\eureka$ like any other action from $\Act$.
To apply test $T$ to process $P$ we form the process \mbox{$T \parT P$}
in which \emph{all} visible actions of $P$ must synchronise with $T$,
and define a set of testing outcomes
$\Apply(T,P)$ where each outcome, in $[0,1]$, arises from a resolution
of the nondeterministic choices in \mbox{$T \parT P$} and gives the
probability that this resolution will reach a \emph{success state},
one in which $\omega$ is possible.

To this end, we inductively define a \emph{results-gathering} function
$\resultsN : \sCSPw \rightarrow \subs{[0,1]}$; it extends to type $\dist{\sCSPw}
\rightarrow \subs{[0,1]}$ via the convention $\results{\Delta} \Defs
\Exp{\Delta}\resultsN$.
\[\begin{array}{@{}r@{~\Defs~}l@{}}
  \results{s} &
  \begin{cases}
                 \sset{1} &\text{if $s \ar{\eureka}$}, \\
\bigcup \setof{\results{\Delta}}{s \ar{\alpha} \Delta}
                     &\text{if $s \nar{\eureka}$ but still $s \rightarrow$},\\
                 \sset{0} &\text{if $s\!\not\rightarrow{}$}
  \end{cases}
\end{array}\]
In the first case above $s \ar{\eureka}$ signifies that $s$ is a success state.
In the second case we mean that $\eureka$ is not possible from
$s$---hence $s$ is not a success state---but that at least one
``non-success'' action $\alpha\in \Act_\tau$ is---and possibly
several---and then the union is over all such $\alpha$.
This is done so that $\resultsN$ accounts for success actions in processes generally;
when applied to test outcomes, however, the only non-success action is $\tau$.
Note that $\resultsN$ is well defined when applied to finite,
loop-free processes, such as the ones of \pCSP.

\begin{defi}\label{d1659}
For any \pCSP process $P$ and test $T$, define
$$\Apply(T,P) ~\Defs~ \resultsN{\Op{T\mathbin{\parT}P}}~.$$
\end{defi}

\noindent
With this definition, the general testing framework of \cite{DNH84}
yields two testing preorders for \pCSP, one based on \emph{may}
testing, written $P \pMayleq Q$, and the other on \emph{must} testing,
written $P \pMustleq Q$.

\begin{defi}\label{d1700}
The \emph{may-} and \emph{must} preorders are given by
\[
 \begin{array}{@{}r@{\hspace{.7em}}c@{\hspace{.7em}}l@{}}
  P \pMayleq Q &\textrm{\rm iff}& \text{for all
  tests $T$: } \Apply(T,P) \Hleq \Apply(T,Q) \\
  P \pMustleq Q &\textrm{\rm iff}& \text{for all
  tests $T$: } \Apply(T,P) \Sleq \Apply(T,Q)
 \end{array}
\]
with $\Hleq,\Sleq$ the Hoare, Smyth preorders on $\subsN{
[0,1]}$.
These are defined as follows:
\begin{align*}
  X \Hleq Y &\;\;\text{iff}\qquad  \forall
  x\inp X\!\! : \exists y\inp Y\!\! : x \leq y\\
X \Sleq Y &\;\;\text{iff}\qquad \forall y\inp Y \!\!: \exists x\inp X\!\! : x \leq y
\end{align*}
\end{defi}

\noindent
In other words, $Q$ is a correct refinement of $P$ in the
probabilistic may-testing preorder if each outcome (in [0,1]) of
applying a test to process $P$ can be matched or increased by applying
the same test to process $Q$.  Likewise, $Q$ is a correct refinement
of $P$ in the probabilistic must-testing preorder if each outcome of
applying a test to $Q$ matches or increases an outcome obtainable by
applying the same test to $P$.

\begin{exa}\label{ex:2}
Consider the test
\[T := a.((b.d.\eureka \probc{\frac{1}{2}} c.e.\eureka) \intc
(b.f.\eureka\probc{\frac{1}{2}} c.g.\eureka))\]
which is graphically depicted in Figure~\ref{fig:examples} (iii).
If we apply $T$ to processes $P$ and $Q$ given in Example~\ref{ex:1},
we form the two processes described in Figure~\ref{fig:test}. It is then
easy to calculate the testing outcomes:
\[\begin{array}{rcl}
\Apply(T,P) & = & \frac{1}{2}\cdot\sset{1,0}
                 +\frac{1}{2}\cdot\sset{1,0} \\
  & = & \sset{0,\ \frac{1}{2},\ 1}\\
\\
\Apply(T,Q) & = & \frac{1}{2}\cdot\sset{\frac{1}{2}}
                 + \frac{1}{2}\cdot\sset{\frac{1}{2}}\\
  & = & \sset{\frac{1}{2}}.
\end{array}\]
We can see that $P$ and $Q$ can be distinguished by the test $T$
since $\Apply(T,P) \not\Hleq \Apply(T,Q)$
and $\Apply(T,Q) \not\Sleq \Apply(T,P)$. In other words, we have $P
\not\pMayleq Q$ and $Q\not\pMustleq P$ because of the witness test $T$.
\end{exa}

\begin{figure}
\vspace{-.7em}
\[\begin{array}{ccc}
\psset{arrows=->}
\pstree[nodesep=0pt,levelsep=20pt,labelsep=.05]{\Tdot}
 { \pstree{\Tdot[dotstyle=o] \tlput{\tau}}
             {\pstree{\Tdot
                 \tlput[tpos=.1]{\scriptscriptstyle{\frac{1}{2}}}}
                 {\pstree{\Tdot[dotstyle=o] \tlput{\tau}}
                    {\pstree{\Tdot \tlput[tpos=.1]
                       {\scriptscriptstyle{\frac{1}{2}}}}
                        {\pstree{\Tdot \tlput{\tau}}
                          {\pstree{\Tdot \tlput{\tau}}
                           {\pstree{\Tdot \tlput{\omega}}{}}
                          }
                        }
                     \pstree{\Tdot \trput[tpos=.1]
                       {\scriptscriptstyle{\frac{1}{2}}}}
                        {\pstree{\Tdot \tlput{\tau}}
                          {\pstree{\Tdot \tlput{\tau}}
                           {\pstree{\Tdot \tlput{\omega}}{}}
                          }
                        }
                    }
                  \pstree{\Tdot[dotstyle=o] \trput{\tau}}
                    {\pstree{\Tdot
                             \tlput[tpos=.1]{\scriptscriptstyle{\frac{1}{2}}}
                            }
                      {\pstree{\Tdot \tlput{\tau}}{}}
                     \pstree{\Tdot
                             \trput[tpos=.1]{\scriptscriptstyle{\frac{1}{2}}}
                            }
                      {\pstree{\Tdot \tlput{\tau}}{}}
                    }
                 }
              \pstree{\Tdot
                 \trput[tpos=.1]{\scriptscriptstyle{\frac{1}{2}}}}
                 {\pstree{\Tdot[dotstyle=o] \tlput{\tau}}
                    {\pstree{\Tdot
                             \tlput[tpos=.1]{\scriptscriptstyle{\frac{1}{2}}}
                            }
                      {\pstree{\Tdot \tlput{\tau}}{}}
                     \pstree{\Tdot
                             \trput[tpos=.1]{\scriptscriptstyle{\frac{1}{2}}}
                            }
                      {\pstree{\Tdot \tlput{\tau}}{}}
                    }
                  \pstree{\Tdot[dotstyle=o] \trput{\tau}}
                    {\pstree{\Tdot \tlput[tpos=.1]
                       {\scriptscriptstyle{\frac{1}{2}}}}
                        {\pstree{\Tdot \tlput{\tau}}
                          {\pstree{\Tdot \tlput{\tau}}
                           {\pstree{\Tdot \tlput{\omega}}{}}
                          }
                        }
                     \pstree{\Tdot \trput[tpos=.1]
                       {\scriptscriptstyle{\frac{1}{2}}}}
                        {\pstree{\Tdot \tlput{\tau}}
                          {\pstree{\Tdot \tlput{\tau}}
                           {\pstree{\Tdot \tlput{\omega}}{}}
                          }
                        }
                    }
                 }
             }
 }

& \qquad\qquad\qquad &
\psset{arrows=->}
\pstree[nodesep=0pt,levelsep=20pt,labelsep=.05]{\Tdot}
 { \pstree{\Tdot[dotstyle=o] \tlput{\tau}}
             {\pstree{\Tdot
                 \tlput[tpos=.1]{\scriptscriptstyle{\frac{1}{2}}}}
                 {\pstree{\Tdot[dotstyle=o] \tlput{\tau}}
                    {\pstree{\Tdot \tlput[tpos=.1]
                       {\scriptscriptstyle{\frac{1}{2}}}}
                        {\pstree{\Tdot \tlput{\tau}}
                          {\pstree{\Tdot \tlput{\tau}}
                           {\pstree{\Tdot \tlput{\omega}}{}}
                          }
                        }
                     \pstree{\Tdot \trput[tpos=.1]
                       {\scriptscriptstyle{\frac{1}{2}}}}
                        {\pstree{\Tdot \tlput{\tau}}
                          {}
                        }
                    }
                  \pstree{\Tdot[dotstyle=o] \trput{\tau}}
                    {\pstree{\Tdot
                             \tlput[tpos=.1]{\scriptscriptstyle{\frac{1}{2}}}
                            }
                      {\pstree{\Tdot \tlput{\tau}}{}}
                     \pstree{\Tdot
                             \trput[tpos=.1]{\scriptscriptstyle{\frac{1}{2}}}
                            }
                      {\pstree{\Tdot \tlput{\tau}}
                        {\pstree{\Tdot \tlput{\tau}}
                           {\pstree{\Tdot \tlput{\omega}}{}}
                        }
                      }
                    }
                 }
              \pstree{\Tdot
                 \trput[tpos=.1]{\scriptscriptstyle{\frac{1}{2}}}}
                 {\pstree{\Tdot[dotstyle=o] \tlput{\tau}}
                    {\pstree{\Tdot
                             \tlput[tpos=.1]{\scriptscriptstyle{\frac{1}{2}}}
                            }
                      {\pstree{\Tdot \tlput{\tau}}{}}
                     \pstree{\Tdot
                             \trput[tpos=.1]{\scriptscriptstyle{\frac{1}{2}}}
                            }
                      {\pstree{\Tdot \tlput{\tau}}
                        {\pstree{\Tdot \tlput{\tau}}
                           {\pstree{\Tdot \tlput{\omega}}{}}
                        }
                      }
                    }
                  \pstree{\Tdot[dotstyle=o] \trput{\tau}}
                    {\pstree{\Tdot \tlput[tpos=.1]
                       {\scriptscriptstyle{\frac{1}{2}}}}
                        {\pstree{\Tdot \tlput{\tau}}
                          {\pstree{\Tdot \tlput{\tau}}
                           {\pstree{\Tdot \tlput{\omega}}{}}
                          }
                        }
                     \pstree{\Tdot \trput[tpos=.1]
                       {\scriptscriptstyle{\frac{1}{2}}}}
                        {\pstree{\Tdot \tlput{\tau}}
                          {}
                        }
                    }
                 }
             }
 }
\\

(i)\ T \parT P
& &
(ii)\  T \parT Q
\end{array}\]
\caption{Testing $P$ and $Q$ with $T$.}\label{fig:test}
\end{figure}

\noindent
In \cite{DGHMZ07} we applied the testing framework described above to
show that many standard laws of CSP are no longer valid in the
probabilistic setting of \pCSP, and to provide counterexamples for a
few distributive laws involving probabilistic choice that may appear
plausible at first sight. We also showed that $P \pMustleq Q$ implies
$Q \pMayleq P$ for all \pCSP processes $P$ and $Q$, i.e.\ that must
testing is more discriminating than may testing and that the preorders
$\pMayleq$ and $\pMustleq$ are oriented in opposite directions.

%-------------------------------------------------------------------------
\section{Simulation and failure simulation}\label{sec:simulations}

\noindent
Let $\mathord{\aRel} \subseteq S \mathop{\times} \dist{S}$ be a
relation from states to distributions.
As in  \cite{DGHMZ07}, we lift it to a relation
$\mathord{\lift{\aRel}} \subseteq \dist{S} \!\mathop{\times}\! \dist{S}$
by letting $\Delta \!\lift{\aRel}\! \Theta$
whenever there is a finite index set $I$ and $p\inp\dist{I}$ such that
\begin{enumerate}[(i)]

\item
$\Delta = \sum_{i\in I}{p_i \cdot \pdist{s_i}}$~,

\item
For each $i \inp I$ there is a distribution $\Phi_i$ s.t.
$s_i \aRel \Phi_i$~,

\item
$\Theta = \sum_{i \in I} {p_i \cdot \Phi_i}$~.

\end{enumerate}
For functions, the lifting operation can be understood as a Kleisli construction on
a probabilistic power domain \cite{Jones}, and was implicit in the work of Kozen \cite{Koz81}; in our more general setting of relations, it can equivalently be
defined in terms of a distribution on $\aRel$, sometimes called
\emph{weight function} (see e.g.\ \cite{JL91,Seg95}).
An important point here is that in the
decomposition (i) of $\Delta_1$ into $\sum_{i\in I}{p_i \cdot
\pdist{s_i}}$, the states $s_i$ are \emph{not necessarily distinct}:
that is, the decomposition is not in general unique. For notational
convenience, the lifted versions of the transition relations
$\ar\alpha$ for $\alpha\inp\Act_\tau$ are again denoted $\ar\alpha$.

We write $s \hatar{\tau} \Delta$ if either $s \ar{\tau}
\Delta$ or $\Delta = \pdist{s}$; again
\mbox{$\Delta_1 \hatar{\tau} \Delta_2$} denotes the lifted relation.
Thus for example we have
$\Op{(a \intc b) \probc{\frac{1}{2}} (a \intc c)}
  \;\hatar{\tau}\; \Op{a \probc{\frac{1}{2}} ((a\intc b)
    \probc{\frac{1}{2}} c)}$
because
  \begin{enumerate}[(i)]\itemsep 1ex
  \item
$\Op{(a \intc b) \probc{\frac{1}{2}} (a \intc c)} =
\frac{1}{4} \cdot \Op{(a \intc b)} + \frac{1}{4} \cdot \Op{(a \intc b)} +
\frac{1}{4} \cdot \Op{(a \intc c)} + \frac{1}{4} \cdot \Op{(a \intc c)} $\;,

\item
\hfill$\begin{array}[t]{cl}
     \Op{(a \intc b)} &\ar{\tau} \Op{a} \\
     \Op{(a \intc b)} &\hatar{\tau} \Op{a \intc b} \\
     \Op{(a \intc c)} &\ar{\tau} \Op{a} \\
     \Op{(a \intc c)} &\ar{\tau} \Op{c} \\
  \end{array}$\hfill\mbox{}

\item
and
$\Op{a \probc{\frac{1}{2}} ((a\intc b)
    \probc{\frac{1}{2}} c)}  =
\frac{1}{4} \cdot \Op{a} + \frac{1}{4} \cdot \Op{(a \intc b)} +
\frac{1}{4} \cdot \Op{a} + \frac{1}{4} \cdot \Op{ c} $\;.

  \end{enumerate}

We now define the weak transition relation $\dar{\hat\tau}$ as the
transitive and reflexive closure $\ar{\hat\tau}^*$ of $\ar{\hat\tau}$,
while for $a\neq \tau$ we let $\Delta_1 \dar{\hat{a}} \Delta_2$ denote
$\Delta_1 \dar{\hat\tau}\ar{a}\dar{\hat\tau} \Delta_2$.
Finally, we write $s\nar{X}$ with $X\subseteq\Act$ when $\forall \alpha\in
X\cup\sset{\tau}: s\nar{\alpha}$, and $\Delta\nar{X}$ when $\forall
s\in\support{\Delta}: s\nar{X}$.
The main properties of the lifted weak transition relations which are used throughout the paper
are given in the following lemma.
\begin{lem}\label{lem:sum}  Suppose $\sum_{i \in I} p_i = 1$ and $\Delta_i \dar{\hat \alpha} \Phi_i$
  for each $i \inp I$, with $I$ a finite index set. Then\vspace{-4pt}
$$\displaystyle\sum_{i \in I}{p_i \cdot \Delta_i} \;\dar{\hat \alpha}\;
       \sum_{i \in I}{p_i \cdot \Phi_i}\;.$$
Conversely, if $\sum_{i \in I}{p_i \cdot \Delta_i} \dar{\hat \alpha}
\Phi$ then $\Phi = \sum_{i \in I}{p_i \cdot \Phi_i}$ for some $\Phi_i$
such that $\Delta_i \dar{\hat \alpha} \Phi_i$ for each $i \inp I$.
\end{lem}

\begin{proof}
The first claim occurs as Lemma 6.6 of \cite{DGHMZ07}.
The second follows by repeated application of Proposition 6.1(ii) of \cite{DGHMZ07},
taking $\aRel$ to be $\hatar{\tau}$ and $\ar{a}$ for $a\inp\Act$.
\end{proof}

\begin{defi}\label{def:failsim}
A relation $\mathord{\aRel} \subseteq \sCSP \mathop{\times} \dist{\sCSP}$ is
said to be a
 \emph{failure simulation} if for all $s,\Theta,\alpha,\Delta,X$ we have
 that
\begin{enumerate}[$\bullet$]

\item $s \aRel \Theta \land s \ar{\alpha}  \Delta$ implies
 $\exists \Theta'\!:\Theta \dar{\hat{\alpha}} \Theta' \land
  \Delta \lift{\aRel} \Theta'$

\item $s \aRel \Theta \land s \nar{X}$ implies
 $\exists \Theta'\!:\Theta \dar{\hat{\tau}} \Theta' \land
  \Theta' \nar{X}$.
\end{enumerate}
We write $s \failsim \Theta$ to mean that there is some failure
simulation $\aRel$ such that $s \aRel \Theta$. Similarly, we define
\emph{simulation}\footnote{It is called \emph{forward
simulation} in \cite{Seg95}.} and $s \forsim \Theta$ by dropping
the second clause in \Def{def:failsim}.\footnote{We have reversed
the orientation of the
  symbols \plat{$\forsiminv$ and $\failsiminv$} w.r.t.\ \cite{DGHMZ07} and
  \cite{DGHMZ07a}; the pointy side now points to a single state, and
  the flat side to a distribution.}
\end{defi}

\begin{defi}\label{def:simulation preorder}
The \emph{simulation preorder} $\forsimpreo$ and \emph{failure simulation
preorder} $\failsimpreo$ on \pCSP are defined as follows:
\[
 \begin{array}{@{}rcl@{}}
  P \forsimpreo Q &\textrm{\rm iff}& \Op{Q} \dar{\hat{\tau}} \Theta
  \text{ for some $\Theta$ with } \Op{P} \liftforsim \Theta \\
  P \failsimpreo Q &\textrm{\rm iff}& \Op{P} \dar{\hat{\tau}} \Theta
  \text{ for some $\Theta$ with } \Op{Q} \liftfailsim \Theta\,.
 \end{array}
\]
(Note the opposing directions.)
The equivalences generated by $\forsimpreo$ and $\failsimpreo$ are
called \emph{(failure) simulation equivalence}, denoted
$\forsimequiv$ and $\failsimequiv$, respectively.
\end{defi}

\begin{exa}
Compare the processes $P = a\probc{\frac{1}{2}}b$ and $P\intc P$.
Note that $\Op{P}$ is the distribution \plat{$\frac{1}{2}\cdot\lift{a} + \frac{1}{2}\cdot\lift{b}$}
whereas $\Op{P \intc P}$ is the point distribution \plat{$\lift{P \intc P}$}.
The relation $\aRel$ given by
  \begin{align*}\textstyle
   (P \intc P)  ~\aRel~ (\frac{1}{2}\cdot\lift{a} + \frac{1}{2}\cdot\lift{b}) \qquad
   a ~\aRel~ \pdist{a} \qquad
   b ~\aRel~ \pdist{b} \qquad
   \Cnil ~\aRel~ \pdist{\Cnil}
  \end{align*}
is a simulation, because the $\tau$-step ${P \intc P}\ar{\tau}
(\frac{1}{2}\cdot\lift{a} + \frac{1}{2}\cdot\lift{b})$ can be matched by the
idle transition $(\frac{1}{2}\cdot\lift{a} + \frac{1}{2}\cdot\lift{b})
\dar{\hat\tau}(\frac{1}{2}\cdot\lift{a} + \frac{1}{2}\cdot\lift{b})$,
and we have $(\frac{1}{2}\cdot\lift{a} + \frac{1}{2}\cdot\lift{b})\lift\aRel
(\frac{1}{2}\cdot\lift{a} + \frac{1}{2}\cdot\lift{b})$.
Thus $(P \intc P) \forsim (\frac{1}{2}\cdot\lift{a} +
\frac{1}{2}\cdot\lift{b}) = \Op{P}$, hence $\mathord{\Op{P \intc P}}
\lift\forsim \Op{P}$, and therefore $\mathord{P \intc P} \forsimpreo P$.

This type of reasoning does not apply to the other direction.
Any simulation $\aRel$ with $(\frac{1}{2}\cdot\lift{a} +
\frac{1}{2}\cdot\lift{b}) \lift\aRel \mathord{\lift{P \intc P}}$ would have
to satisfy $a \aRel \mathord{\lift{P \intc P}}$ and $b \aRel \mathord{\lift{P \intc P}}$.
However, the move $a \ar{a}\Cnil$ cannot be matched by the process
$\lift{P \intc P}$, as the only transition the latter process can do
is $\lift{P \intc P} \ar{\tau} (\frac{1}{2}\cdot\lift{a} +
\frac{1}{2}\cdot\lift{b})$, and only half of that distribution can
match the $a$-move. Thus, no such simulation exists, and we
find $\mathord{\Op{P}} \not\lift\forsim \Op{P \intc P}$.
Nevertheless, we still have $\mathord{P} \forsimpreo P \intc P$.
Here, the transition $\dar{\hat\tau}$ from Definition~\ref{def:simulation preorder}
comes to the rescue. As $\Op{P \intc P} \dar{\hat\tau} \Op{P}$ and
$\Op{P} \lift\forsim \Op{P}$, we obtain $\mathord{P} \forsimpreo P \intc P$.
\end{exa}

\begin{exa}
Let $P=a \probc{\frac{1}{2}} b$ and $Q = P\extc P$. We have $P\forsimpreo Q$ because
$\Op{P}\lift{\forsim}\Op{Q}$ which comes from the following observations:
\begin{enumerate}[(1)]
\item $\Op{P}=\frac{1}{2}\cdot\lift{a} + \frac{1}{2}\cdot\lift{b}$
\item $\Op{Q}=\frac{1}{2}\cdot (\frac{1}{2}\cdot\lift{a\extc a}+ \frac{1}{2}\cdot\lift{a\extc b})
  + \frac{1}{2}\cdot (\frac{1}{2}\cdot\lift{b\extc a}+\frac{1}{2}\cdot\lift{b\extc b})$
\item $a\forsim (\frac{1}{2}\cdot\lift{a\extc a}+\frac{1}{2}\cdot\lift{a\extc b})$
\item $b\forsim (\frac{1}{2}\cdot\lift{b\extc a}+\frac{1}{2}\cdot\lift{b\extc b})$
\end{enumerate}
This kind of reasoning does not apply to $\failsim$. For example, we
have $a\not\failsim (\frac{1}{2}\cdot\lift{a\extc a}+\frac{1}{2}\cdot\lift{a\extc
  b})$ because the state on the left hand side can refuse to do action
$b$ while the distribution on the right hand side cannot. Indeed, it
holds that $Q\not\failsimpreo P$.
\end{exa}

We have already shown in \cite{{DGHMZ07}} that $\forsimpreo$ is a
precongruence and that it implies $\pMayleq$. Similar results
can be established for $\failsimpreo$ as well.
Below we summarise these facts.
\begin{prop}\label{prp:cong}
Suppose $\mathord{\sqsubseteq}\in\sset{\forsimpreo, \failsimpreo}$.
Then $\sqsubseteq$ is a preorder, and if $P_i \sqsubseteq Q_i$ for $i =
1,2$ then $a.P_1 \sqsubseteq a.Q_1$ for $a\inp\Act$ and
  $P_1 \odot P_2 ~\sqsubseteq~ Q_1 \odot Q_2~$ for
  $\odot\inp\,\{\intc,\,\extc,\,\probc{p},\,\Par{A}\}$.
\end{prop}
\begin{proof}
The case $\forsimpreo$ was proved in
\cite[Corollary~6.10 and Theorem~6.13]{DGHMZ07};
the case $\failsimpreo$ is analogous. As an example, we show that
$\failsimpreo$ is preserved under parallel composition. The key step
is to show that the binary relation $\mathord{\aRel}\subseteq
  \sCSP\times\dist{\sCSP}$  defined by
  \[\aRel ~:=~ \sset{(s_1|_A s_2, \Delta_1|_A \Delta_2)\mid
    s_1\failsim\Delta_1 \wedge s_2\failsim\Delta_2}.\]
   is a failure simulation.

Suppose $s_i\failsim\Delta_i$ for $i=1,2$ and $s_1\Par{A} s_2\nar{X}$
for some $X\subseteq\Act$. For each
$a\in X$ there are two possibilities:
\begin{enumerate}[$\bullet$]
\item If $a\not\in A$ then $s_1\nar{a}$ and $s_2\nar{a}$, since
  otherwise we would have $s_1\Par{A} s_2\ar{a}$.
\item If $a\in A$ then either $s_1\nar{a}$ or $s_2\nar{a}$,
  since otherwise we would have $s_1\Par{A} s_2\ar{\tau}$.
\end{enumerate}
Hence we can partition the set $X$ into three subsets: $X_0$, $X_1$
and $X_2$ such that $X_0=X\backslash A$ and $X_1\cup X_2\subseteq A$
with $s_1\nar{X_1}$ and $s_2\nar{X_2}$, but allowing $s_1\nar{a}$ for
some $a\in X_2$ and $s_2\nar{a}$ for some $a\in X_1$.
We then have that $s_i\nar{X_0\cup X_i}$ for $i=1,2$.
By the assumption that $s_i\failsim\Delta_i$ for
$i=1,2$, there is a $\Delta'_i$ with
$\Delta_i\dar{\hat{\tau}}\Delta'_i\nar{X_0\cup X_i}$.
Therefore $\Delta'_1|_A\Delta'_2\nar{X}$ as well.
It is stated in \cite[Lemma 6.12(i)]{DGHMZ07} that
if $\Phi \dar{\hat\tau} \Phi'$ then
$\Phi \Par{A} \Delta \dar{\hat\tau} \Phi'\Par{A} \Delta$ and
$\Delta \Par{A} \Phi \dar{\hat\tau} \Delta \Par{A} \Phi'$.
So we have
$\Delta_1\Par{A}\Delta_2\dar{\hat{\tau}}\Delta'_1\Par{A}\Delta'_2$.
Hence $\Delta_1\Par{A}\Delta_2$ can match up the failures of $s_1\Par{A} s_2$.

The matching up of transitions and the using of $\aRel$ to prove the
preservation property of $\failsimpreo$ under parallel composition are
similar to those in the corresponding proof
for simulations \cite[Theorem 6.13(v)]{DGHMZ07}, so we omit them.
\end{proof}

\noindent
We recall the following result from \cite[Theorem 6.17]{DGHMZ07}.

\begin{thm}\label{thm:forsim.sound}
If $P \forsimpreo Q$ then $P \pMayleq Q$.
\end{thm}

\begin{proof}
For any test $T\in \pCSPw$ and process $P\in\pCSP$ the set
$\results{T\parT P}$ is finite, so
\begin{equation}\label{max.i}
P \pMayleq Q \textrm{ iff }
\max(\results{\Op{T\parT P}}) \leq \max(\results{\Op{T\parT Q}})
\textrm{ for every test $T$.}
\end{equation}
The following properties for $\Delta_1,\Delta_2\in\pCSPw$ and
$\alpha\in\Act_\tau$ are not hard to establish:
\begin{equation}\label{max.ii}
\Delta_1 \dar{\hat{\alpha}} \Delta_2 \textrm{ implies }
\max(\results{\Delta_1}) \geq \max(\results{\Delta_2}).
\end{equation}
\begin{equation}\label{max.iii}
\Delta_1 \lift{\forsim} \Delta_2 \textrm{ implies }
\max(\results{\Delta_1}) \leq \max(\results{\Delta_2}).
\end{equation}
In \cite[Lemma~6.15 and Proposition~6.16]{DGHMZ07} similar properties
are proven using a function {\it maxlive} instead of $\max\!\circ\resultsN$.
The same arguments apply here.

Now suppose $P\forsimpreo Q$. Since $\forsimpreo$ is preserved by the
parallel operator we have that $T\parT P \forsimpreo T\parT Q$ for an
arbitrary test $T$. By definition, this means that there is a
distribution $\Delta$ such that $\Op{T\parT Q}\dar{\hat{\tau}}\Delta$ and
$\Op{T\parT P}\lift{\forsim}\Delta$. By (\ref{max.ii}) and (\ref{max.iii}) we infer that
$\max(\results{\Op{T\parT P}}) \leq \max(\results{\Op{T\parT Q}})$.
The result now follows from (\ref{max.i}).
\end{proof}

\noindent
It is tempting to use the same idea to prove that $\failsimpreo$
implies $\pMustleq$, but now using the function $\min\!\circ\resultsN$.
However, the $\min$-analogue of Property (\ref{max.ii}) is in general invalid.
For example, let $R$ be the process $a\parT (a\extc\eureka)$. We have $\min(\results{R})=1$,
yet $R\ar{\tau}\pdist{\Cnil\parT\Cnil}$ and $\min(\results{\pdist{\Cnil\parT\Cnil}})=0$.
Therefore, it is not the case that $\Delta_1 \dar{\hat{\tau}} \Delta_2$ implies
$\min(\results{\Delta_1}) \leq \min(\results{\Delta_2})$.

Our strategy is therefore as follows.  Write
$s\ar{\alpha}_\eureka\Delta$ if both $s\nar{\eureka}$ and
$s\ar{\alpha}\Delta$ hold. We define $\ar{\hat{\tau}}_\eureka$ as
$\ar{\hat{\tau}}$ using $\ar{\tau}_\eureka$ in place of
$\ar{\tau}$. Similarly we define $\dar{}_\eureka$ and
$\dar{\hat{\alpha}}_\eureka$.  Thus the subscript $\eureka$ on a
transition of any kind indicates that no state is passed through in
which $\eureka$ is enabled.  A version of failure simulation adapted
to these transition relations is then defined as follows.

\begin{defi}
Let $\mathord{\failsime}\subseteq \sCSPw\times\dist{\sCSPw}$ be the largest
relation such that $s\failsime\Theta$ implies
\begin{enumerate}[$\bullet$]
\item if $s\ar{\alpha}_\eureka\Delta$ then there is some $\Theta'$
  with $\Theta\dar{\hat{\alpha}}_\eureka\Theta'$ and $\Delta\lift{\failsime}\Theta'$
\item if $s\nar{X}$ with $\eureka\in X$ then there is some $\Theta'$
  with $\Theta\dar{\hat{\tau}}_\eureka\Theta'$ and $\Theta'\ar{X}$.
\end{enumerate}
Let $P \failsimpreo^{e} Q$ iff $\Op{P} \dar{\hat{\tau}}_\eureka \Theta$
for some $\Theta$ with $\Op{Q} \lift{\failsime} \Theta$.
\end{defi}

\noindent
Note that for processes $P,Q$ in \pCSP (as opposed to $\pCSP^\eureka$), we have
$P\failsimpreo Q$ iff $P\failsimpreo^e Q$.

\begin{prop}\label{prop:failecongruence}
If $P,Q$ are processes in \pCSP with $P\failsimpreo Q$ and $T$ is
a process in $\pCSP^\eureka$ then $T\parT P\failsimpreo^e T\parT Q$.
\end{prop}
\begin{proof}
Similar to the proof of Proposition~\ref{prp:cong}.
\end{proof}

\begin{prop}\label{prop:presound}
The following properties hold for $\min\!\circ\resultsN$,
with $\Delta_1,\Delta_2\in \dist{\sCSPw}$:
\begin{equation}\label{min.i}
  P \pMustleq Q \textrm{ \rm iff }
  \min(\results{\Op{T\parT P}}) \leq \min(\results{\Op{T\parT Q}})
  \textrm{ \rm for every test $T$.}
\end{equation}
\begin{equation}\label{min.ii}
  \Delta_1 \dar{\hat{\alpha}}_\eureka \Delta_2
  \textrm{ \rm for $\alpha\in\Act_\tau$ implies }
  \min(\results{\Delta_1}) \leq \min(\results{\Delta_2}).
\end{equation}
\begin{equation}\label{min.iii}
  \Delta_1\lift{\failsime}\Delta_2 \textrm{ \rm implies }
  \min(\results{\Delta_1}) \geq \min(\results{\Delta_2}).
\end{equation}
\end{prop}
\noindent\textit{Proof.\/}
Property (\ref{min.i}) is again straightforward, and Property (\ref{min.ii}) can be
established just as in Lemma~6.15 in \cite{DGHMZ07}, but with all
$\leq$-signs reversed. Property (\ref{min.iii}) follows by structural induction,
simultaneously with the property, for $s\in\sCSPw$ and $\Delta\in\dist{\sCSPw}$, that
\begin{equation}\label{min.iv}
s \failsime \Delta \mbox{~~implies~~} \min(\results{s}) \geq \min(\results{\Delta})\;.
\end{equation}
The reduction of Property (\ref{min.iii}) to (\ref{min.iv}) proceeds exactly
as in \cite[Lemma~6.16(ii)]{DGHMZ07}. For (\ref{min.iv}) itself
we distinguish three cases:
\begin{enumerate}[$\bullet$]
\item If $s\ar{\eureka}$, then $\min(\results{s})=1\geq\min(\results{\Delta})$ trivially.
\item If $s\nar{\eureka}$ but $s\rightarrow$, then we can closely
  follow the proof of \cite[Lemma~6.16(i)]{DGHMZ07}:\\
  Whenever $s \ar{\alpha}_\omega \Theta$, for $\alpha\in\Act_\tau$ and
  $\Theta\in\dist{\sCSPw}$, then $s \failsime \Delta$ implies the
  existence of some $\Delta_\Theta$ such that
  $\Delta\ar{\hat{\alpha}}_\omega^*\Delta_\Theta$ and $\Theta
  \lift{\failsime} \Delta_\Theta$. By induction, using (\ref{min.iii}),
  it follows that $\min(\results{\Theta}) \geq \min(\results{\Delta_\Theta})$.
  Consequently, we have that
  \[\begin{array}{@{~~~~}rcll}
  \min(\results{s}) & = & \min(\sset{\min(\results{\Theta}) ~|~ s \ar{\alpha} \Theta})\\
            & \geq  & \min(\sset{\min(\results{\Delta_\Theta}) ~|~ s \ar{\alpha} \Theta}) \\
            & \geq  & \min(\sset{\min(\results{\Delta}) ~|~ s \ar{\alpha} \Theta}) &
                                                \qquad\mbox{(by (\ref{min.ii}))}\\
                 & = & \min(\results{\Delta})\;.
  \end{array}\]
\item If $s\not\rightarrow$, that is $s\nar{\Act^\omega}$, then
  there is some $\Delta'$ such that $\Delta\dar{\hat{\tau}}_\eureka\Delta'$
  and $\Delta'\nar{\Act^\omega}$. By the definition of
  $\resultsN$, $\min(\results{\Delta'})\mathbin=0$.
  Using (\ref{min.ii}), we have $\min(\results{\Delta})\leq\min(\results{\Delta'})$,
  so $\min(\results{\Delta})\mathbin=0$ as well.
  Thus, also in this case $\min(\results{s}) \geq \min(\results{\Delta})$.
\hfill\qed
\end{enumerate}

\begin{thm}\label{t:simtest}
If $P \failsimpreo Q$ then $P \pMustleq Q$.
\end{thm}
\begin{proof}
Similar to the proof of Theorem~\ref{thm:forsim.sound}, using (\ref{min.i})--(\ref{min.iii}).
\end{proof}

\noindent
The next four sections are devoted to proving the converse of
Theorems~\ref{thm:forsim.sound} and~\ref{t:simtest}.

%-------------------------------------------------------------------------
\section{State- versus action-based testing}\label{sec:state-vs-action}

\noindent
Much work on testing \cite{DNH84,WL92,DGHMZ07} uses success
\emph{states} marked by outgoing $\eureka$-actions;
this is referred to as \emph{state-based} testing, which we have used in
\Sect{sec:testing} to define the preorders $\Mayleq$ and $\Mustleq$.
In other work
\cite{Seg96,DGMZ07}, however, it is the \emph{actual execution} of
$\eureka$ that constitutes success.
This \emph{action-based} approach is formalised
as in the state-based approach,  via a modified results-gathering function:

\[\begin{array}{@{}r@{\,\Defs\,}l@{}}
\Dresults{s}
  &
  \begin{cases}
\bigcup \setof{\!\Dresults{\Delta}\!}{\!s \Ar{\alpha} \Delta {\,\land\, \alpha\neq\eureka}} \cup
        \setof{\!1\!}{\!s \Ar{\eureka}}
                     &\text{if $s\!\rightarrow$}\\
                 \sset{0} &\text{\hspace{-1.4em}otherwise}
  \end{cases}
\end{array}\]
As in the original $\resultsN$, the $\alpha$'s are non-success
actions, including $\tau$; and again, this is done for generality,
since in testing outcomes the only non-success action is $\tau$.

If we use this results-gathering function rather than $\resultsN{}$ in
Definitions~\ref{d1659} and \ref{d1700} we obtain the two slightly
different testing preorders, $ \pdMayleq$ and $\pdMustleq$.  The
following proposition shows that state-based testing is at least as
discriminating as action-based testing:
\vfill\eject

\begin{prop}\label{p:action-based}\mbox{}
\begin{enumerate}[\em(1)]

\item If $P \pMayleq Q$ then $P \pdMayleq Q$.

\item If $P \pMustleq Q$ then $P \pdMustleq Q$.

\end{enumerate}
\end{prop}
\begin{proof}
For any action-based test $\Overline{T}$ we construct a
state-based test $T$ by replacing each subterm $\omega.Q$ by
$\tau.\omega$; then we have $\resultsN{\Op{T\parT P}} =
  \DresultsN{\Op{\Overline{T}\parT P}}$ for all \pCSP processes $P$.
\end{proof}

\noindent
\Prop{p:action-based} enables us to reduce our main goal,
the converse of Theorems~\ref{thm:forsim.sound} and~\ref{t:simtest}, to the following property.
\begin{thm}\label{t:simtest-reverse}\mbox{}
\begin{enumerate}[\em(1)]

\item If $P \pdMayleq Q$ then $P \forsimpreo Q$.

\item If $P \pdMustleq Q$ then $P \failsimpreo Q$.
\end{enumerate}
\end{thm}

\noindent
We set the proof of this theorem as our goal in the next three sections.
\pagebreak[2]

Once we have obtained this theorem, it follows that in our
framework of finite probabilistic processes the state-based and
action-based testing preorders coincide. This result no longer holds
in the presence of divergence, at least for must-testing.
\begin{exa}\label{ex:div}
Suppose we extend our syntax with a state-based process $\Omega$, to model divergence,
and the operational semantics of Figure~\ref{fig:opsem} with the rule
\begin{align*}
  \Omega \ar{\tau} \pdist{\Omega}.
\end{align*}
It is possible to extend the results-gathering functions $\resultsN$
and $\DresultsN$ to these infinite processes, although the definitions
are no longer inductive (cf.\ Definition 5 of \cite{DGMZ07} or
  Definition~\ref{def:results} of the appendix).
In this extended setting we will have
$a.\Omega \not \pMustleq a. \Omega \intc 0$ because of the test $a.\omega$:\vspace{-1em}
\begin{align*}
\results{\Op{a.\omega \mathbin{\parT} a.\Omega}} = \sset{1}
\mbox{~~~ while  ~~~}
\results{\Op{a.\omega \mathbin{\parT} a.\Omega \intc 0}} = \sset{0,1}.
\end{align*}
This intuitively is due to the fact that the $\Omega$-encoded
divergence of the left-hand process occurs only \emph{after} the first
action $a$; and since the left-hand process cannot deadlock before
that action, relation $\Mustleq$ would prevent the right-hand process
from doing so.

However, a peculiarity of action-based testing is that success
actions can be indefinitely inhibited by infinite $\tau$-branches.
We have
\begin{align*}
  \Dresults{\Op{a.\omega \mathbin{\parT} a.\Omega}} =
\Dresults{\Op{a.\omega \mathbin{\parT} a.\Omega \intc 0}} = \sset{0,1}.
\end{align*}
Indeed no test can be found to distinguish them, and so one can show
$a.\Omega  \pdMustleq a.\Omega \intc 0$.
\end{exa}

\noindent
Note that probabilistic behaviour plays no role in this counter-example.
In CSP (without probabilities) there is no difference between
$\dMayleq$ and $\Mayleq$, whereas $\dMustleq$ is strictly less
discriminating than $\Mustleq$. For finitely branching processes, the
CSP refinement preorder based on failures and divergences
\cite{BHR84,csp,OH86} coincides with the state-based relation $\Mustleq$.

%-------------------------------------------------------------------------
\section{Vector-based testing}\label{sec:vector}

\noindent
This section describes another variation on testing,
a richer testing framework due to Segala
\cite{Seg96}, in which countably many success actions exist: the
application of a test to a process yields a set of \emph{vectors} over
the real numbers, rather than a set of scalars. The resulting
action-based testing preorders will serve as a stepping stone in proving
\Thm{t:simtest-reverse}.

Let $\Omega$ be a \emph{set} of fresh success actions with
$\Omega\cap\Act_\tau=\emptyset$. An $\Omega$-test is again a \pCSP
process, but this time allowing subterms $\omega.P$ for any
$\omega\inp\Omega$.  Applying such a test to a process yields a non-empty
set of test outcome-\emph{tuples} $\MDApply(T,P) \subseteq [0,1]^\Omega$.
As with standard scalar testing, each outcome arises from a
resolution of the nondeterministic choices in \mbox{$T \parT P$}.
However, here an  outcome is a  \emph{tuple} and its $\omega$-component gives
the probability that this resolution will perform the success action $\omega$.

For vector-based testing  we again inductively define a
results-gathering function, but first we require some auxiliary notation.
For any action $\alpha$ define $\alpha!:[0,1]^\Omega \rightarrow
[0,1]^\Omega$ by
\begin{align*}
 \alpha!o(\omega)  =
  \begin{cases}
    1  &\text{if  $\omega\Eq\alpha$}\\
    o(\omega) &\text{otherwise}
  \end{cases}
\end{align*}
so that if $\alpha$ is a success action, in $\Omega$, then $\alpha!$
updates the tuple to 1 at that point, leaving it unchanged otherwise,
and when $\alpha\not\in\Omega$ the function $\alpha!$ is the identity.
These functions lift to sets $O\subseteq[0,1]^\Omega$ as usual, via
$\alpha!O\Defs\{\alpha!o\mid o\inp O\}$.

Next, for any set $X$ define its
\emph{convex closure} $\CClose X$ by
\[\textstyle
\CClose X \Defs \setof{\sum_{i\in I}p_i o_i }{p\inp\dist{I}
    \mbox{ and } o: I\rightarrow X}~ .
\]
Here, as usual, $I$ is assumed to be a finite index set.
Finally, $\vec0 \in [0,1]^\Omega$ is given by $\vec0(\omega)=0$ for all
$\omega\in\Omega$. Let \pCSPO be the set of $\Omega$-tests, and
\sCSPO the set of state-based $\Omega$-tests.

\begin{defi}\label{def:vector-testing}
The \emph{action-based, vector-based, convex-closed results-gathering function}
$\MDCresultsN : \sCSPO \rightarrow \subs{[0,1]^\Omega}$ is given by
\begin{equation}\label{e:def.vec}
\begin{array}{@{}r@{~~\Defs~~}l@{}}
\MDCresults{s} &
\begin{cases}
\CClose \bigcup \setof{\alpha!(\MDCresults{\Delta})}{s \Ar{\alpha}
  \Delta,\ \alpha\in\Omega\cup{\Act_\tau}}
                     &\text{if $s\rightarrow$}\\
                 \sset{\vec{0}} &\text{\hspace{-1.4em}otherwise}
\end{cases}
\end{array}
\end{equation}
As with our previous results-gathering functions
$\resultsN$ and $\DresultsN$, this function extends to the type
\plat{$\dist{\sCSPO} \rightarrow \subs{[0,1]^\Omega}$} via the convention
$\MDCresults{\Delta} \Defs \Exp{\Delta}\MDCresultsN$.

For any \pCSP process $P$ and $\Omega$-test $T$, let
$$\MDCApply(T,P) ~\Defs~ \MDCresultsN{\Op{T \mathbin{\parT} P}}~.$$
The \emph{vector-based} \emph{may-} and \emph{must} preorders are given by
\[
 \begin{array}{@{}r@{\hspace{.7em}}c@{\hspace{.7em}}l@{}}
  P \pdmMayleq Q &\textrm{\rm iff}& \text{for all $\Omega$-tests $T$: }
                  \MDCApply(T,P) \Hleq \MDCApply(T,Q) \\
  P \pdmMustleq Q &\textrm{\rm iff}& \text{for all $\Omega$-tests $T$: }
                   \MDCApply(T,P) \Sleq \MDCApply(T,Q)
 \end{array}
\]
where $\Hleq$ and $\Sleq$ are the Hoare- and Smyth preorders on
$\subsN{[0,1]^\Omega}$ generated from $\leq$ index-wise on $[0,1]^\Omega$ itself.
\end{defi}

\noindent
We will explain the r\^ole of convex-closure $\CClose$ in this
definition. Let $\MDresultsN$ be defined as \plat{$\MDCresultsN$}
above, but omitting the use of $\CClose$. It is easy to see that
\plat{$\MDCresultsN(s) = \CClose \MDresultsN(s)$} for all $s\in \sCSPO$.

Applying convex closure to subsets of the one-dimensional interval
$[0,1]$ (such as arise from applying scalar tests to processes) has no
effect on the Hoare and Smyth orders between these subsets:

\begin{lem}\label{lem:cc}
  Suppose $X,\; Y \subseteq [0,1]$. Then
  \begin{enumerate}[\em(1)]
  \item $X \Hleq Y$ if and only if $\CClose X \Hleq \CClose Y$.
  \item $X \Sleq Y$ if and only if $\CClose X \Sleq \CClose Y$.
  \end{enumerate}
\end{lem}
\begin{proof}
We restrict attention to $(1)$; the proof of $(2)$ goes likewise.
 It suffices to show that (i) $X \Hleq
  \CClose X$ and (ii) $\CClose X \Hleq X$. We only prove (ii) since
  (i) is obvious. Suppose $x\in \CClose X$, then $x=\sum_{i\in I}p_i x_i$ for
  a finite set $I$ with $\sum_{i\in I}p_i=1$ and $x_i\in X$. Let $x^*=
  \max\sset{x_i\mid i\in I}$. Then
\\\mbox{}\hfill$\displaystyle
  x~=~\sum_{i\in I}p_i x_i ~\leq~ \sum_{i\in I}p_i x^* ~=~ x^* ~\in~ X.
  $
\end{proof}

\noindent
It follows that for scalar testing it makes no difference whether
convex closure is employed or not. Vector-based testing, as proposed in
Definition~\ref{def:vector-testing}, is a conservative extension of
action-based testing, as described in \Sect{sec:state-vs-action}:
\begin{cor}
  Suppose $\Omega$ is the singleton set $\sset{\eureka}$. Then
  \begin{enumerate}[\em(1)]
  \item $ P \pdmMayleq Q$ if and only if $P \pdMayleq Q$.
  \item $ P \pdmMustleq Q$ if and only if $P \pdMustleq Q$.
  \end{enumerate}
\end{cor}
\begin{proof}
$\MDCresultsN = \CClose\MDresultsN = \CClose\DresultsN$ when $\Omega$ is $\sset{\omega}$,
so the result follows from \Lem{lem:cc}.
\end{proof}

\noindent
\Lem{lem:cc} does not generalise to $[0,1]^k$, when $k > 1$, as the following example
demonstrates:
\begin{exa}\label{ex:ccnot}
Let $X,\;Y$ denote $\sset{(0.5, 0.5)},\; \sset{(1,0), (0,1)}$
respectively. Then it is easy to show that $\CClose X \Hleq \CClose Y$
although obviously $X \not\Hleq Y$.
\end{exa}

\noindent
This example can be exploited to show that for vector-based testing it
\emph{does} make a difference whether convex closure is employed.

\begin{exa}\label{ex:convex}
Consider the two processes
\[P:= a \probc{\frac{1}{2}} b \quad
 {\rm \ and\ } \quad Q := a \intc b ~.\]
Take $\Omega=\sset{\omega_1,\omega_2}$. Employing the
results-gathering function $\MDresultsN$, without convex closure,
with the test  $T:=  a.\omega_1 \extc b.\omega_2$ we obtain
\begin{align*}
\MDApply(T, P) & ~~~=~~~ \sset{
 (0.5, 0.5)}\\
 \MDApply(T, Q) & ~~~=~~~ \sset{(1,0), (0,1)} ~.
\end{align*}
As pointed out in Example~\ref{ex:ccnot}, this entails
$\MDApply(T, P) \not\Hleq \MDApply(T, Q)$, although their convex
closures $\MDCApply(T, P)$ and $\MDCApply(T, Q)$ \emph{are} related under
the Hoare preorder.
\end{exa}

\noindent
Convex closure is a uniform way of ensuring that internal choice can
simulate an arbitrary probabilistic choice \cite{HSM97}. For the
processes $P$ and $Q$ of Example~\ref{ex:convex} it is obvious that $P
\forsimpreo Q$, and from \Thm{thm:forsim.sound} it therefore follows that $P \pMayleq Q$.
This fits with the intuition that a probabilistic choice is an
acceptable implementation of a nondeterministic choice occurring in a
specification. Considering that we use $\pdmMayleq$ as a stepping
stone in showing the coincidence of $\forsimpreo$ and $\pMayleq$, we
must have $P \pdmMayleq Q$. For this reason we use convex closure in \Def{def:vector-testing}.

In \cite{DGMZ07} the results-gathering function $\MDCresultsN$
with $\Omega= \{\omega_1,\omega_2,\cdots\}$ was called simply $\mathbb W$ (because
action-based/vector-based/convex-closed testing was assumed there throughout, making the
$\widehat{\cdot}_{\scriptscriptstyle\updownarrow}^{\raisebox{-.2em}{\scriptsize$~\Omega$}}$-indicators
superfluous); and it was defined in terms of a formalisation of the notion of a resolution.
As we show in Proposition~\ref{p1126} of the appendix,
the  inductive Definition~\ref{def:vector-testing} above yields the same
results.  In the present paper our interest in vector-based testing stems
from the following result.

\begin{thm}\label{t:multiUni}\mbox{}
\begin{enumerate}[\em(1)]
 \item $P \pdmMayleq Q$ iff $P \pdMayleq Q$
 \item $P \pdmMustleq Q$ iff $P \pdMustleq Q$.
\end{enumerate}
\end{thm}

\begin{proof}
In \cite[Theorem 3]{DGMZ07} this theorem has been established for versions
of \plat{$\pdmMayleq$} and \plat{$\pdmMustleq$} where tests are finite
probabilistic automata, as defined in our Appendix~\ref{app:resolutions}.
The key argument is that when $P \pdmMayleq Q$ can be refuted by
means of a vector-based test $T$, then \plat{$P \pdMayleq Q$} can be refuted
by means of a scalar test $T\|U$, where U is administrative code
which collates the vector of results produced by $T$ and effectively
renders them as a unique scalar result, and similarly for \plat{$\pdmMustleq$}.
This theorem applies to our setting as well, due to the observation
that if a test $T$ can be represented as a $\pCSPO$-expression, then
so can the test $T\|U$.
\end{proof}

\noindent
Because of  \Thm{t:multiUni}, in order to establish
\Thm{t:simtest-reverse} it will suffice to show that
\begin{enumerate}[(1)]
\item $P \pdmMayleq Q$ implies $P \forsimpreo Q$ and

\item $P \pdmMustleq Q$ implies $P \failsimpreo Q$.
\end{enumerate}
This shift from scalar testing to vector-based testing is motivated by the fact that
the latter enables us to use more informative tests, allowing us to discover
more intensional properties of the processes being tested.

The crucial characteristics of $\MDCApply$ needed for the above implications
are summarised in Lemmas~\ref{lem:inductive} and \ref{l:inductive.5}.
For convenience of presentation, we write
$\vec{\eureka}$ for the vector in $[0,1]^\Omega$  defined by
$\vec{\eureka}(\eureka) = 1$ and $\vec{\eureka}(\eureka') = 0$ for
$\eureka'\neq\eureka$. Sometimes we treat a distribution
$\Delta$ of finite support as the \pCSP expression $\bigoplus_{s\in
  \support{\Delta}}\Delta(s)\Cdot s$, so that $\MDCApply(T,\Delta)
\Defs \Exp{\Delta}\MDCApply(T, \_\!\_)$.

\begin{lem}\label{lem:inductive}
Let $P$ be a \pCSP process, and $T,T_i$ be tests.
\begin{enumerate}[\em(1)]
\item
  $\outc \in \MDCApply(\omega,P)$ iff $\outc=\vec{\eureka}$.
\item $\vec0\in\MDCApply(\bigbox_{a\in X}a.\omega,P)$ iff
  $\exists\Delta:\Op{P}\dar{\hat\tau}\Delta\nar{X}$.
\item
  Suppose the action $\omega$ does not occur in the test $T$.
  Then $\outc\in\MDCApply(\omega \mathop\Box a.T,P)$ with $o(\omega)=0$ iff
  there is a $\Delta\inp \dist{\sCSP}$
  with $\Op{P}\dar{\hat a}\Delta$ and \plat{$\outc\in\MDCApply(T,\Delta)$}.
\item $\outc\in\MDCApply(\bigoplus_{i\in I} p_i\Cdot T_i,P)$ iff
  $\outc=\sum_{i\in I}p_i\outc_i$ for some $\outc_i\in\MDCApply(T_i,P)$.
\item $\outc\in\MDCApply(\bigsqcap_{i\in I}\!T_i,P)$ if for all $i\inp I$
  there are $q_i\inp\,[0,1]$ and $\Delta_i\inp\dist{\sCSP}$ such that
  $\sum_{i\in I}q_i=1$, $\Op{P}\dar{\hat\tau}\sum_{i\in I}q_i\cdot\Delta_i$
  and $\outc=\sum_{i\in I}q_i\outc_i$ for some
  \plat{$\outc_i\in\MDCApply(T_i,\Delta_i)$}.
\end{enumerate}
\end{lem}
\begin{proof}
Straightforward, by induction on the structure of $P$.
\end{proof}

\noindent
The converse of Lemma~\ref{lem:inductive} (5) also holds, as the
following lemma says. However, the proof is less straightforward.

\begin{lem}\label{l:inductive.5}
Let $P$ be a \pCSP process, and $T_i$ be tests.
If $\outc\in\MDCApply(\bigsqcap_{i\in I}\!T_i,P)$ then for all $i\inp I$
  there are $q_i\inp\,[0,1]$ and $\Delta_i\inp\dist{\sCSP}$ with
  $\sum_{i\in I}q_i = 1$ such that
  $\Op{P}\dar{\hat\tau}\sum_{i\in I}q_i\cdot\Delta_i$ and
  $\outc=\sum_{i\in I}q_i\outc_i$ for some
  \plat{$\outc_i\inp\MDCApply(T_i,\Delta_i)$}.
\end{lem}

\noindent\textit{Proof.\/}
Given that the states of our pLTS are \sCSP expressions, there exists a
well-founded order on the combination of states in \sCSP and distributions in
$\dist{\sCSP}$, such that $s \ar{\alpha} \Delta$ implies that $s$ is larger than
$\Delta$, and any distribution is larger than the states in its
support. Intuitively, this order corresponds to the usual order on
natural numbers if we graphically depict a pLTS as a finite tree
(cf.~Section~\ref{sec:pcsp}) and assign to each node a number to indicate
its level in the tree.
Let $T=\bigsqcap_{i\in I}\!T_i$.
We prove the following two claims
\begin{enumerate}[(a)]
\item
If $s$ is a state-based process and $o\in\MDCApply(T,s)$ then there
are some $\sset{q_i}_{i\in I}$ with  $\sum_{i\in I}q_i=1$ such that
$s \dar{\hat{\tau}}
\sum_{i\in I} q_i \cdot \Delta_i$, $o=\sum_{i\in I} q_i  o_i$,
and \plat{$o_i \in \MDCApply(T_i,\Delta_i)$}.

\item
If $\Delta \in \dist{\sCSP}$ and $o\in\MDCApply(T,\Delta)$ then there
are some $\sset{q_i}_{i\in   I}$ with  $\sum_{i\in I}q_i=1$ such that
$\Delta \dar{\hat{\tau}}
\sum_{i\in I} q_i \cdot \Delta_i$, $o=\sum_{i\in I} q_i  o_i$,
and \plat{$o_i \in \MDCApply(T_i,\Delta_i)$}.
\end{enumerate}
by simultaneous induction on the
order mentioned above, applied to $s$ and $\Delta$.

\begin{enumerate}[(a)]
\item
 We have two sub-cases depending on whether $s$ can make an initial $\tau$-move
 or not.
 \begin{enumerate}[$\bullet$]
 \item
  If $s$ cannot make a $\tau$-move, that is $s\nar{\tau}$, then the only
  possible moves from $T\parT s$ are $\tau$-moves originating in $T$;
  $T$ has no non-$\tau$ moves, and any non-$\tau$ moves that
  might be possible for $s$ on its own are inhibited by the alphabet
  $\Act$ of the composition. Suppose $o\in\MDCApply(T,s)$. Then by
  definition (\ref{e:def.vec}) there are some $\sset{q_i}_{i\in I}$
  with $\sum_{i\in I}q_i=1$ such that $o=\sum_{i\in I}q_i o_i$ and
  \plat{$o_i \in\MDCApply(T_i,s) = \MDCApply(T_i,\pdist{s})$}.
  Obviously we also have $\Op{s}\dar{\hat{\tau}} \sum_{i\in I}q_i \cdot \pdist{s}$.

 \item
  If $s$ can make one or more $\tau$-moves, then we have
  $s \ar{\tau} \Delta'_j$ for $j\in J$, where without loss of
  generality $J$ can be assumed to be a non-empty
  finite set disjoint from $I$, the index set for $T$.
  The possible first moves for $T\parT s$ are $\tau$-moves either of
  $T$ or of $s$, because $T$ cannot make initial non-$\tau$ moves and that
  prevents a proper synchronisation from occurring on the first step.
  Suppose that $o \in \MDCApply(T,s)$. Then
  by definition (\ref{e:def.vec}) there are some $\sset{p_k}_{k\in
  I\cup J}$ with
  $\sum_{k\in I\cup J} p_k = 1$ and
  \begin{align}
   \label{eq:1a}
   o = \sum_{k\in I\cup J} p_k  o'_k \\
   \label{eq:1b}
   o'_i \in \MDCApply(T_i,s) \qquad & \text{for all $i\in I$} \\
   o'_j \in \MDCApply(T,\Delta_j) \qquad & \text{for all $j\in J$.}
  \end{align}
  For each $j\in J$, we know by the induction hypothesis that
  \begin{align}
   \label{eq:2a}
   \Delta'_j \dar{\hat{\tau}} \sum_{i\in I}p_{ji} \cdot \Delta'_{ji}
   \\
   \label{eq:2b}
   o'_j = \sum_{i\in I}p_{ji}  o'_{ji} \\
   \label{eq:2c}
   o'_{ji} \in \MDCApply(T_i,\Delta'_{ji})
  \end{align}
  for some $\sset{p_{ji}}_{i\in I}$ with $\sum_{i\in I}p_{ji} = 1$.
  Let
  \begin{align*}
   q_i & = p_i + \sum_{j\in J}p_j p_{ji}\\
   \Delta_i & = \frac{1}{q_i}(p_i \cdot \pdist{s} + \sum_{j\in J}p_j
   p_{ji} \cdot \Delta'_{ji}) \\
   o_i & = \frac{1}{q_i}(p_i  o'_i + \sum_{j\in J}p_j
   p_{ji}  o'_{ji})
  \end{align*}
  for each $i\in I$, except that $\Delta_i$ and $o_i$ are chosen
    arbitrarily in case $q_i=0$. It can be checked by arithmetic that $q_i,
  \Delta_i, o_i$ have the required properties, viz.~that $\sum_{i\in
  I} q_i = 1$, that $o = \sum_{i\in I}q_i o_i$ and that
  \begin{align*}
   s & \dar{\hat{\tau}}  \sum_{i\in I}p_i\cdot\pdist{s} + \sum_{j\in
   J}p_j \cdot \Delta'_j \\
   & \dar{\hat{\tau}}  \sum_{i\in I}p_i\cdot\pdist{s} + \sum_{j\in
   J}p_j \cdot \sum_{i\in I}p_{ji}\cdot \Delta'_{ji} \qquad\text{by
   (\ref{eq:2a}) and Lemma~\ref{lem:sum}}\\
   & =  \sum_{i\in I}q_i\cdot\Delta_i \;.
  \end{align*}
 Finally, it follows from (\ref{eq:1b}) and (\ref{eq:2c}) that $o_i
 \in \MDCApply(T_i,\Delta_i)$ for each $i\in I$.
 \end{enumerate}

\item
 Let $\support{\Delta} = \sset{s_j}_{j\in J}$ and
 $r_j=\Delta(s_j)$. W.l.o.g.\ we may assume that
 $J$ is a non-empty finite set disjoint from $I$.
 Using that $\MDCApply(T,\Delta) \Defs \Exp{\Delta}\MDCApply(T, \_\!\_)$,
 if $o\in\MDCApply(T,\Delta)$ then\vspace{-1em}
 \begin{align}
 \label{eq:4a}
 o=\sum_{j\in J}r_j  o'_j \\
 o'_j \in \MDCApply(T,s_j)
 \end{align}
 For each $j\in J$, we know by the induction hypothesis that
 \begin{align}
 \label{eq:5a}
 s_j \dar{\hat{\tau}} \sum_{i\in I}q_{ji} \cdot \Delta'_{ji}\\
 \label{eq:5b}
 o'_j = \sum_{i\in I}q_{ji}  o'_{ji}\\
 \label{eq:5c}
 o'_{ji} \in \MDCApply(T_i,\Delta'_{ji})
 \end{align}
 for some $\sset{q_{ji}}_{i\in I}$ with $\sum_{i\in I}q_{ji}=1$. Thus let
 \begin{align*}
  q_i & = \sum_{j\in J}r_j q_{ji}\\
  \Delta_i & =  \frac{1}{q_i}\sum_{j\in J}r_j q_{ji} \cdot
  \Delta'_{ji}\\
  o_i & =  \frac{1}{q_i}\sum_{j\in J}r_j q_{ji}
  o'_{ji}
 \end{align*}
again choosing $\Delta_i$ and $o_i$ arbitrarily in case $q_i=0$.
As in the first case, it can be shown by arithmetic that the
collection $r_i, \Delta_i, o_i$ has the required properties.
\qed\end{enumerate}

%-------------------------------------------------------------------------
\section{Modal logic}\label{sec:modal-logic}

\noindent
  In this section we present logical characterisations $\modalSpreo$
  and $\modalFSpreo$ of our testing preorders. Besides their intrinsic
  interest, these logical preorders also serves as a stepping stone in
  proving Theorem~\ref{t:simtest-reverse}. In this section we show
  that the logical preorders are sound w.r.t.\ the simulation and
  failure simulation preorders, and hence w.r.t.\ the testing
  preorders; in the next section we establish completeness.
To start, we define a set $\FSL$ of modal formulae, inductively, as follows:
\begin{enumerate}[$\bullet$]
\item $\refuse{X} \in \FSL$ when $X\subseteq \Act$,
\item $\diam{a}\phi \in\FSL$ when $\phi\inp \FSL$ and $a\inp \Act$,
\item $\bigwedge_{i\in I}\phi_i\in\FSL$ when $\phi_i \inp \FSL$ for all
  $i\inp I$, with $I$ finite,
\item and $\bigoplus_{i\in I}p_i\cdot\phi_i \in \FSL$ when $p_i\inp [0,1]$ and
  $\phi_i \inp \FSL$ for all $i\inp I$, with $I$ a finite index set,
  and $\sum_{i\in I}p_i=1$.
\end{enumerate}
We often write $\phi_1 \land \phi_2$ for $\bigwedge_{i\in \{1,2\}}\phi_i$
and $\top$ for $\bigwedge_{i\in \emptyset}\phi_i$.

The \emph{satisfaction relation} $\models\:\subseteq \dist{\sCSP} \times \FSL$ is given by:
\begin{enumerate}[$\bullet$]
\item $\Delta\models \refuse{X}$ iff there
  is a $\Delta'$ with $\Delta \dar{\hat\tau}\Delta'$ and \plat{$\Delta'\nar{X}$},
\item $\Delta\models \diam{a}\phi$ iff there is a $\Delta'$
  with $\Delta \dar{\hat a}\Delta'$ and $\Delta'\models \phi$,
\item $\Delta\models\bigwedge_{i\in I}\phi_i$ iff
  $\Delta\models\phi_i$ for all $i\inp I$
\item and $\Delta\models\bigoplus_{i\in I}p_i\cdot\phi_i$ iff there are $\Delta_i\!\in \dist{\sCSP}$, for all $i\inp I$,
with $\Delta_i\models \phi_i$,
  such that $\Delta \dar{\hat\tau}\sum_{i\in I}p_i\cdot\Delta_i$.
\end{enumerate}
Let $\SL$ be the subclass of $\FSL$ obtained by skipping the $\refuse{X}$ clause.
We write $P \modalSpreo Q$ just when $\Op{P}\models\phi$ implies
$\Op{Q}\models\phi$ for all $\phi\inp \SL$, and $P\modalFSpreo Q$
just when $\Op{P}\models\phi$ is implied by $\Op{Q}\models\phi$ for
all $\phi\inp \FSL$. (Note the opposing directions.)

In order to obtain the main result of this section,
Theorem~\ref{thm-characteristic formulae}, we introduce the following tool.
\begin{defi}\label{d:char.formula}
The \emph{$\FSL\!$-characteristic formula} $\phi\hspace{-.8pt}_s$ or
$\phi\hspace{-.8pt}_\Delta$ of a process $s\inp\sCSP$ or
$\Delta\inp\dist{\hspace{-.2pt}\sCSP\hspace{-.2pt}}$ is defined inductively:
\begin{enumerate}[$\bullet$]
\item $\phi_s \Defs \bigwedge_{s \ar{a} \Delta} \diam{a}\phi_\Delta \land \refuse{\{a \mid s \nar{a}\}}$
       if $s\nar{\tau}$,
\item $\phi_s \Defs \bigwedge_{s \ar{a} \Delta} \diam{a}\phi_\Delta \land
       \bigwedge_{s \ar{\tau} \Delta} \phi_\Delta$ otherwise,
\item $\phi_\Delta \Defs \bigoplus_{s \in \support\Delta} \Delta(s)\cdot\phi_s$.
\end{enumerate}
Here the conjunctions $\bigwedge_{s \ar{a} \Delta}$ range over suitable pairs $a,\Delta$, and $\bigwedge_{s \ar{\tau} \Delta}$ ranges over suitable $\Delta$.
The \emph{$\SL$-characteristic formulae} $\psi_s$ and $\psi_\Delta$ are defined likewise, but
omitting the conjuncts $\refuse{\{a \mid s \nar{a}\}}$.
\end{defi}

\noindent
Write $\phi\Rrightarrow \psi$ with $\phi,\psi\inp\FSL$ if for each
  distribution $\Delta$ one has $\Delta \models \phi$ implies $\Delta \models \psi$.
  Then it is easy to see that $\phi_{\overline{s}}\Lleftarrow\!\!\!\Rrightarrow \phi_s$
  and $\bigwedge_{i\in I}\phi_i \Rrightarrow \phi_i$ for any $i\inp
  I$; furthermore, the following property can be established by an easy inductive proof.
\begin{lem}\label{lem-characteristic formulae}
For any $\Delta\inp\dist\sCSP$ we have
$\Delta \models \phi_\Delta$, as well as $\Delta \models \psi_\Delta$.
\hfill\qed
\end{lem}

\noindent
It and the following lemma help to establish \Thm{thm-characteristic formulae}.

\begin{lem}\label{characteristic formulae}
For any processes $P,Q\in\pCSP$ we have that
$\Op{P} \models \phi_{\Ops{Q}}$ implies $P \failsimpreo Q$, and likewise
that $\Op{Q} \models \psi_{\Ops{P}}$ implies $P \forsimpreo Q$.
\end{lem}
\begin{proof}
To establish the first statement, we
define the relation $\aRel$ by $s \aRel \Theta$ iff $\Theta \models
\phi_s$; to show that it is a failure simulation we first prove the
following technical result:
\begin{equation}\label{eq:modal}
\Theta \models \phi_\Delta ~~\mbox{implies}~~ \exists \Theta':
\Theta\dar{\hat\tau}\Theta' \land \Delta\mathrel{\overline\aRel}\Theta'.
\end{equation}
Suppose $\Theta \models \phi_\Delta$
with $\phi_\Delta = \bigoplus_{i\in I}p_i\cdot\phi_{s_i}$,
so that we have $\Delta = \sum_{i\in I}p_i\cdot\pdist{s_i}$ and
for all $i\inp I$ there are $\Theta_i\inp \dist{\sCSP}$ with
$\Theta_i\models \phi_{s_i}$ such that
$\Theta \dar{\hat\tau}\Theta'$ with
\plat{$\Theta'\Defs\sum_{i\in I}p_i\cdot\Theta_i$}.
Since $s_i\aRel\Theta_i$ for all $i\inp I$ we have
$\Delta\mathrel{\overline\aRel}\Theta'$.

Now we show that $\aRel$ is a failure simulation.
\begin{enumerate}[$\bullet$]
\item Suppose ${s} \aRel \Theta$ and $s
  \ar{\tau} \Delta$. Then from Definition~\ref{d:char.formula} we have
 $\phi_{s} \Rrightarrow \phi_\Delta$, so that
  $\Theta \models \phi_{\Delta}$. Applying (\ref{eq:modal}) gives us
  $\Theta\dar{\hat\tau}\Theta'$ with
  $\Delta\mathrel{\overline\aRel}\Theta'$ for some $\Theta'$.
\item Suppose $s \aRel \Theta$ and $s
  \ar{a} \Delta$ with $a\inp \Act$. Then $\phi_{s}
  \Rrightarrow \diam{a}\phi_\Delta$, so $\Theta \models
  \diam{a}\phi_{\Delta}$. Hence $\exists\Theta'$ with $\Theta
  \dar{\hat a} \Theta'$ and $\Theta' \models \phi_{\Delta}$. Again apply
  (\ref{eq:modal}).
\item Suppose ${s} \aRel \Theta$ and
  \plat{$s\nar{X}$} with $X\subseteq A$. Then $\phi_{{s}}
  \Rrightarrow \refuse{X}$, so $\Theta\models\refuse{X}$.
  Hence $\exists \Theta'$ with $\Theta \dar{\hat\tau}
  \Theta'$ and \plat{$\Theta'\nar{X}$}.
\end{enumerate}
Thus $\aRel$ is indeed a failure simulation. By our assumption
$\Op{P} \models \phi_{\Ops{Q}}$, using \Eqn{eq:modal}, there
exists a $\Theta'$ such that $\Op{P}\dar{\hat{\tau}}\Theta'$ and $\Op{Q}
\mathrel{\overline\aRel} \Theta'$, which gives
$P \failsimpreo Q$ via \Def{def:simulation preorder}.

To establish the second statement,
define the relation $\aRelS$ by $s \aRelS \Theta$ iff $\Theta \models
\psi_s$; exactly as above one obtains
\begin{equation}\label{eq:modalS}
\Theta \models \psi_\Delta ~~\mbox{implies}~~ \exists \Theta':
\Theta\dar{\hat\tau}\Theta' \land \Delta\mathrel{\overline\aRelS}\Theta'.
\end{equation}
Just as above it follows that $\aRelS$ is a simulation.
By the assumption $\Op{Q} \models \phi_{\Ops{P}}$, using \Eqn{eq:modalS}, there
exists a $\Theta'$ such that $\Op{Q}\dar{\hat{\tau}}\Theta'$ and $\Op{P}
\mathrel{\overline\aRelS} \Theta'$. Hence
$P \forsimpreo Q$ via \Def{def:simulation preorder}.
\end{proof}

\begin{thm}\label{thm-characteristic formulae}\mbox{}
\begin{enumerate}[\em(1)]

\item If $P \modalSpreo Q$ then $P \forsimpreo Q$.

\item If $P \modalFSpreo Q$ then $P \failsimpreo Q$.
\end{enumerate}
\end{thm}

\begin{proof}
Suppose $P\modalFSpreo Q$. By \Lem{lem-characteristic formulae}
we have $\Op{Q}\!\models\!\phi_{\Ops{Q}}$ and hence $\Op{P}\models\phi_{\Ops{Q}}$.
\Lem{characteristic formulae} gives $P\failsimpreo Q$.

For (1), assuming $P\modalSpreo Q$, we have
$\Op{P}\models\psi_{\Ops{P}}$, hence
$\Op{Q}\models\psi_{\Ops{P}}$, and thus $P \forsimpreo Q$.
\end{proof}

%-------------------------------------------------------------------------
\section{Characteristic tests}\label{sec:chartests}

\noindent
Our final step towards \Thm{t:simtest-reverse} is taken in this
section, where we show that every modal formula $\phi$ can
be characterised by a vector-based test $T_\phi$ with the property that any \pCSP
process satisfies $\phi$ just when it passes the test $T_\phi$.

\begin{lem}\label{lem:char tests}
For every $\phi\inp\FSL$ there exists a pair $(T_\phi,v_\phi)$ with
$T_\phi$ an $\Omega$-test and $v_\phi \in [0,1]^\Omega$, such that\vspace{-7pt}
\begin{equation}\label{eq:chartest}
\Delta\models\phi \quad{\it iff}\quad \exists o \in
\MDCApply({T_\phi,\Delta}):~o\leq v_\phi
\vspace{3pt}
\end{equation}
for all $\Delta\inp\dist{\sCSP}$, and in case $\phi\in\SL$ we also have
\begin{equation}\label{eq:chartestmay}
\Delta\models\phi \quad{\it iff}\quad \exists o \in
\MDCApply({T_\phi,\Delta}):~o\geq v_\phi\,.
\end{equation}
\end{lem}
\noindent
$T_\phi$ is called a \emph{characteristic test} of $\phi$ and $v_\phi$
its \emph{target value}.

\begin{proof}
First of all note that if a pair $(T_\phi,v_\phi)$ satisfies the
requirements above, then  any pair obtained from $(T_\phi,v_\phi)$
by bijectively renaming the elements of $\Omega$ also satisfies these
requirements. Hence a characteristic test can always be chosen in such
a way that there is a success action $\omega\inp\Omega$ that does not
occur in (the finite) $T_\phi$. Moreover, any countable collection of characteristic
tests can be assumed to be \emph{$\Omega$-disjoint}, meaning that no
$\omega\inp\Omega$ occurs in two different elements of the collection.

The required characteristic tests and target values are obtained as
follows.
\begin{enumerate}[$\bullet$]
\item Let $\phi = \top$. Take $T_\phi \Defs \omega$ for some
  $\omega\inp\Omega$, and $v_\phi\!\Defs\vec\omega$.
\item Let $\phi=\refuse X$ with $X\subseteq\Act$.
  Take $T_\phi \Defs \bigbox_{a\in X}a.\omega$ for some $\omega\inp\Omega$,
  and $v_\phi \Defs {\vec 0}$.
\item Let $\phi=\diam{a}\psi$. By induction, $\psi$
  has a characteristic test $T_\psi$ with target value $v_\psi$.
  Take $T_\phi \Defs \omega \mathop\Box a.T_\psi$ where
  $\omega\inp\Omega$ does not occur in $T_{\psi}$,
  and $v_\phi \Defs v_\psi$.
\item Let $\phi=\bigwedge_{i\in I}\phi_i$ with $I$
  a finite and non-empty index set. Choose a $\Omega$-disjoint family
  $(T_i,v_i)_{i\in I}$ of characteristic tests $T_i$ with target values
  $v_i$ for each $\phi_i$.  Furthermore, let $p_i\inp\,(0,1]$ for $i\inp
  I$ be chosen arbitrarily such that $\sum_{i\in I}p_i=1$.  Take
  $T_\phi \Defs \bigoplus_{i\in I}p_i\Cdot T_{i}$ and
  $v_\phi \Defs \sum_{i\in I}p_iv_{i}$.
\item Let $\phi=\bigoplus_{i\in I}p_i\cdot\phi_i$. Choose a $\Omega$-disjoint family
  $(T_i,v_i)_{i\in I}$ of characteristic tests $T_i$ with target values
  $v_i$ for each $\phi_i$, such that there are distinct success actions
  $\omega_i$ for $i\inp I$ that do not occur in any of those tests.
  Let \plat{$T'_i \Defs T_i \probc{\frac{1}{2}} \omega_i$} and
  \plat{$v'_i\Defs\frac{1}{2}v_i + \frac{1}{2}\vec{\omega_i}$}.
  Note that for all $i\inp I$ also $T'_i$ is a characteristic test of
  $\phi_i$ with target value $v'_i$.
  Take $T_\phi \Defs \bigsqcap_{i\in I} T'_{i}$
  and $v_\phi \Defs \sum_{i\in I}p_iv'_i$.
\end{enumerate}
Note that $v_\phi(\omega)=0$ whenever $\omega\inp\Omega$ does not
occur in $T_\phi$.
By induction on $\phi$ we now check (\ref{eq:chartest}) above.
\begin{enumerate}[$\bullet$]
\item Let $\phi=\top$. For all $\Delta\in\dist{\sCSP}$ we have
  $\Delta\models\phi$ as well as $\exists o \in
  \MDCApply({T_\phi,\Delta}):~o\leq v_\phi$, using \Lem{lem:inductive}(1).
\item Let $\phi=\refuse X$ with $X\subseteq\Act$.
Suppose $\Delta\models\phi$. Then there is a $\Delta'$
  with $\Delta \dar{\hat\tau}\Delta'$ and \plat{$\Delta'\nar{X}$}.
By \Lem{lem:inductive}(2), $\vec{0}\inp \MDCApply({T_\phi,\Delta})$.

\noindent
Now suppose $\exists o \inp \MDCApply({T_\phi,\Delta}):~o\leq v_\phi$.
This implies $o=\vec{0}$, so by \Lem{lem:inductive}(2) there is a
$\Delta'$ with $\Delta \dar{\hat\tau}\Delta'$ and \plat{$\Delta'\nar{X}$}.
Hence $\Delta\models\phi$.
\item Let $\phi=\diam{a}\psi$ with $a\inp\Act$.
Suppose $\Delta\models\phi$. Then there is a $\Delta'$
  with $\Delta \dar{\hat a}\Delta'$ and $\Delta'\models \psi$.
By induction, $\exists o \inp \MDCApply({T_\psi,\Delta'}):~o\leq v_\psi$.
By \Lem{lem:inductive}(3), $o\inp \MDCApply({T_\phi,\Delta})$.

\noindent
Now suppose $\exists o \inp \MDCApply({T_\phi,\Delta}):~o\leq v_\phi$.
This implies $o(\omega)=0$, so by \Lem{lem:inductive}(3) there is a
$\Delta'$ with $\Delta \dar{\hat a}\Delta'$ and \plat{$o \inp \MDCApply({T_\psi,\Delta'})$}.
By induction, $\Delta'\!\models\! \psi$, so $\Delta\!\models\!\phi$.

\item Let $\phi=\bigwedge_{i\in I}\phi_i$ with $I$ a
finite and non-empty index set.
Suppose $\Delta\models\phi$. Then $\Delta\models\phi_i$ for all $i\inp I$,
and hence, by induction, \plat{$\exists o_i \inp\, \MDCApply({T_i,\Delta}):~o_i\leq v_i$}.
Thus $o\Defs\sum_{i\in I}p_io_i$ \plat{$\inp \MDCApply({T_\phi,\Delta})$}
by \Lem{lem:inductive}(4), and $o\leq v_\phi$.

\noindent
Now suppose $\exists o \inp \MDCApply({T_\phi,\Delta}):~o\leq v_\phi$.
Then, using \Lem{lem:inductive}(4), $o=\sum_{i\in I}p_io_i$ for
certain \plat{$o_i\inp\MDCApply({T_i,\Delta})$}.
Note that $(T_i)_{i\in I}$ is an $\Omega$-disjoint family of tests.
One has \mbox{$o_i\leq v_i$} for all $i\inp I$, for if $o_i(\omega)> v_i(\omega)$
for some $i\inp I$ and $\omega\inp \Omega$, then $\omega$ must occur
in $T_i$ and hence cannot occur in $T_j$ for $j\mathop{\neq} i$. This implies
$v_j(\omega)=0$ for all $j\neq i$ and thus $o(\omega)>v_\phi(\omega)$,
in contradiction with the assumption.  By induction,
$\Delta\models\phi_i$ for all $i\inp I$, and hence $\Delta\models\phi$.
\item Let $\phi=\bigoplus_{i\in I}p_i\cdot\phi_i$. Suppose $\Delta\models\phi$.
Then for all $i\inp I$ there are $\Delta_i\inp
\dist{\sCSP}$ with $\Delta_i\models \phi_i$ such that $\Delta
\dar{\hat\tau}\sum_{i\in I}p_i\cdot\Delta_i$.
By induction, there are $o_i \inp \MDCApply(T_i,\Delta_i)$ with $o_i\leq v_i$.
Hence, there are $o'_i \inp \MDCApply(T'_i,\Delta_i)$ with $o'_i\leq v'_i$.
Thus $o\Defs\sum_{i\in I}p_io'_i\inp \MDCApply({T_\phi,\Delta})$
by \Lem{lem:inductive}(5), and $o\leq v_\phi$.

\noindent
Now suppose $\exists o \inp \MDCApply({T_\phi,\Delta}):~o\leq v_\phi$.
Then, by \Lem{l:inductive.5}, there are $q\inp\dist{I}$ and $\Delta_i$,
for $i\inp I$, such that \plat{$\Delta\dar{\hat\tau}\sum_{i\in I}q_i\cdot\Delta_i$}
and $o=\sum_{i\in I}q_io'_i$ for some \plat{$o'_i\inp\MDCApply(T'_i,\Delta_i)$}.
Now $\forall i:o'_i(\omega_i)=v'_i(\omega_i)=\frac{1}{2}$,
so, using that $(T_i)_{i\in I}$ is an $\Omega$-disjoint family of tests,
$\frac{1}{2}q_i = q_io'_i(\omega_i) = o(\omega_i) \leq
v_\phi(\omega_i) = p_iv'_i(\omega_i) = \frac{1}{2}p_i$.
As $\sum_{i\in I}q_i = \sum_{i\in I}p_i = 1$, it must be that
$q_i = p_i$ for all $i\inp I$.
Exactly as in the previous case one obtains $o'_i\leq v'_i$ for all $i\inp I$.
Given that $T'_i = T_i \probc{\frac{1}{2}} \omega_i$, using \Lem{lem:inductive}(4),
it must be that $o'=\frac{1}{2}o_i + \frac{1}{2}\vec{\omega_i}$ for some
\plat{$o_i \in \MDCApply(T_i,\Delta_i)$} with $o_i \leq v_i$.
By induction, $\Delta_i\models\phi_i$ for all $i\inp I$, and hence
$\Delta\models\phi$.
\end{enumerate}
In case $\phi\inp\SL$, the formula cannot be of the form $\refuse{X}$. Then
a straightforward induction yields that
$\sum_{\omega\in\Omega}v_\phi(\omega)=1$ and
for all $\Delta \inp \dist{\pCSP}$ and $o\inp
\MDCApply({T_\phi,\Delta})$ we have $\sum_{\omega\in\Omega}o(\omega)=1$.
Therefore, $o\leq v_\phi$ iff $o\geq v_\phi$ iff $o = v_\phi$, yielding
(\ref{eq:chartestmay}).
\end{proof}

\begin{thm}\label{thm-characteristic tests}\mbox{}
\begin{enumerate}[\em(1)]

\item If $P \pdmMayleq Q$ then $P \modalSpreo Q$.

\item If $P \pdmMustleq Q$ then $P \modalFSpreo Q$.
\end{enumerate}
\end{thm}

\begin{proof}
Suppose $P\pdmMustleq Q$ and $\Op{Q}\models\phi$
for some $\phi\inp\FSL$. Let $T_\phi$ be a characteristic test of
$\phi$ with target value $v_\phi$. Then \Lem{lem:char tests}
yields $\exists o \inp \MDCApply({T_\phi,\Op{Q}}): o\leq v_\phi$, and
hence, given that $P\pdmMustleq Q$ and $\MDCApply({T_\phi,\Op{R}})=
\MDCApply({T_\phi,R})$ for any $R\in\pCSP$, by the Smyth preorder
we have $\exists o' \in
\MDCApply({T_\phi,\Op{P}}):~o'\leq v_\phi$. Thus $\Op{P}\models\phi$.

The may-case goes likewise, via the Hoare preorder.
\end{proof}

\noindent
Combining Theorems~\ref{t:multiUni}, \ref{thm-characteristic tests}
and \ref{thm-characteristic formulae},
we obtain
\Thm{t:simtest-reverse}, the goal we set ourselves in \Sect{sec:state-vs-action}.
Thus, with Theorems~\ref{thm:forsim.sound} and~\ref{t:simtest}
and \Prop{p:action-based},  we have shown that the may preorder coincides
with simulation and that the must preorder coincides with failure simulation.
These results also imply the converse of both statements in
Theorem~\ref{thm-characteristic tests}, and thus that the logics
$\SL$ and $\FSL$ give logical characterisations of the simulation
and failure simulation preorders $\forsimpreo$ and $\failsimpreo$.
\hfill{\qed}

%---------------------------------------------------
\section{Equational theories}\label{sec:equations}

\begin{figure}
\renewcommand{\Eeq}{=}
\vspace{-.7em}
  \begin{displaymath}
    \begin{array}[t]{@{}lrcl@{}}
     \Ename{P1}            & P \probc{p} P&\Eeq& P\\
      \Ename{P2}            & P \probc{p} Q&\Eeq& Q \probc{1-p} P\\
      \Ename{P3}            & (P \probc{p} Q) \probc{q} R&\Eeq&
                       P \probc{p\cdot q} (Q \probc{\frac{(1-p)\cdot q}{1-p\cdot q}} R)\\
      \Ename{I1}            & P \intc P&\Eeq& P\\
      \Ename{I2}            & P \intc Q&\Eeq& Q \intc P\\
      \Ename{I3}            & (P \intc Q) \intc R&\Eeq& P \intc (Q \intc R)\\
      \Ename{E1}            & P \extc \Cnil&\Eeq& P\\
      \Ename{E2}            & P \extc Q&\Eeq& Q \extc P\\
      \Ename{E3}            & (P \extc Q) \extc R&\Eeq& P \extc (Q \extc R)\\
      \Ename{EI}            & a.P \extc a.Q&\Eeq& a.P \intc a.Q\\
      \Ename{D1}            & P \extc (Q \probc{p} R)&\Eeq& (P \extc Q) \probc{p} (P \extc R)\\
      \Ename{D2}            &  a.P \extc (Q \intc R)  &\Eeq& (a.P \extc Q) \intc (a.P \extc R) \\
      \Ename{D3}            &(P_1 \intc P_2) \extc (Q_1 \intc Q_2)&\Eeq&
                             (P_1 \extc (Q_1 \intc Q_2)) \intc (P_2 \extc (Q_1 \intc Q_2)) \\
                          &&&\quad\intc ((P_1 \intc P_2) \extc Q_1) \intc ((P_1 \intc P_2) \extc Q_2)\\
   \end{array}\!\!\!\!\!
  \end{displaymath}

  \caption{Common equations}
  \label{fig:cequations}
\end{figure}

\noindent
Having settled the problem of characterising the may preorder in terms
of simulation, and the must preorder in terms of failure simulation,
we now turn to complete axiomatisations of the preorders.

In order to focus on the essentials we consider just those \pCSP
processes that do not use the parallel operator $\Par{A}$; we call
the resulting sub-language \nCSP. For a brief discussion of the
axiomatisation for terms involving $\Par{A}$ and the other parallel
operators commonly used in \CSP see \Sect{sec:conc}.

Let us write $P \Eeq Q$ for equivalences that can be derived using the equations
given in \Fig{fig:cequations}.
Given the way we defined the syntax of \pCSP, axiom \Ename{D1} is merely a
case of abbreviation-expansion; thanks to \Ename{D1} there is no
  need for (meta-)variables ranging over the sub-sort of state-based
  processes anywhere in the axioms.
Many of the standard equations for \CSP \cite{csp} are missing;
they are not sound for $\failsimequiv$.
Typical examples include:
\begin{align*}
  a.(P \intc Q) &= a.P \intc a.Q\\
              P &= P \extc P \\
P \extc (Q \intc R) &= (P \extc Q) \intc (P \extc R)\\
P \intc (Q \extc R) &= (P \intc Q) \extc (P \intc R)
\end{align*}
For a detailed discussion of the standard equations for \CSP in the
presence of probabilistic processes see Section 4 of \cite{DGHMZ07}.
\begin{prop}
  Suppose $P \Eeq Q$. Then $P \failsimequiv Q$.
\end{prop}
\begin{proof}
  Because of \Prop{prp:cong}, that $\failsimpreo$ is a precongruence,
  it is sufficient to
  exhibit witness failure simulations for the axioms in
  \Fig{fig:cequations}.
  These are exactly the same as the witness
  simulations for the same axioms, given in \cite{DGHMZ07}. The only
  axiom for which it is nontrivial to check that these simulations are
  in fact failure simulations is \Ename{EI}. That axiom, as stated in
  \cite{DGHMZ07}, is unsound here; it will return in the next section as
  \Ename{May0}. But the special case of $a=b$ yields the axiom
  \Ename{EI} above, and then the witness simulation from \cite{DGHMZ07}
  is a failure simulation indeed.
\end{proof}
\noindent
As $\forsimequiv$ is a less discriminating equivalence than $\failsimequiv$
it follows that $P \Eeq Q$ implies $P \forsimequiv Q$.

This equational theory allows us to reduce terms to a form in which
the external choice operator is applied to prefix terms only.

\begin{defi}[Normal forms]
The set of \emph{normal forms} $N$ is given by the following grammar:
  \begin{align*}
    N &::= N_1 \probc{p} N_2 \BNFsep N_1 \intc N_2 \BNFsep \bigbox_{i \in I} a_i.N_i
  \end{align*}
\end{defi}

\begin{prop}\label{p:nf}
 For every $P \in \nCSP$ there is a normal form $N$ such that $P \Eeq N$.
\end{prop}

\begin{proof}
A fairly straightforward induction, heavily relying on \Ename{D1}--\Ename{D3}.
\end{proof}

\noindent
We can also show that the axioms \Ename{P1}--\Ename{P3} and \Ename{D1} are
in some sense all that are required to reason about probabilistic choice. Let
$P \Peq Q$ denote that equivalence of $P$ and $Q$ can be derived using
those axioms alone. Then we have the following property.

\begin{lem}\label{lem:prob}
Let $P,Q \in \nCSP$. Then $\Op{P} = \Op{Q}$ implies $P \Peq Q$.
\end{lem}

\noindent
Here $\Op{P} = \Op{Q}$ says that $\Op{P}$ and $\Op{Q}$ are the very
same distributions of state-based processes in $\sCSP$; this is a much
stronger prerequisite than $P$ and $Q$ being testing equivalent.

\begin{proof}
The axioms \Ename{P1}--\Ename{P3} and \Ename{D1} essentially allow any processes to
be written in the unique form $\bigoplus_{i\in I}p_i s_i$, where the
$s_i \in \sCSP$ are all different.
\end{proof}

%-------------------------------------------------------------------------
\section{Inequational theories}\label{sec:inequations}

\begin{figure}
\emph{May:}
  \begin{displaymath}
    \begin{array}[t]{@{}lrl@{}}
      \Ename{May0}            & a.P \extc b.Q &=~ a.P \intc b.Q\\
      \Ename{May1}            &  P &\sqsubseteq~ P \intc Q\\
      \Ename{May2}            &  \Cnil & \sqsubseteq~  P \\
     \Ename{May3}            & a.(P \probc{p} Q)&\sqsubseteq~ a.P \probc{p} a.Q\\
   \end{array}
  \end{displaymath}

\vspace{1em}

\emph{Must:}
\begin{displaymath}
\begin{array}[t]{@{}lrl@{}}
\Ename{Must1}  & P \intc Q  & \sqsubseteq~ Q\\
\Ename{Must2}  & \displaystyle R \intc \bigsqcap_{i\in I}\bigoplus_{j\in J_i}p_j
                   \Cdot (a_i.Q_{ij} \extc P_{ij}) & \displaystyle\sqsubseteq~
                   \bigbox_{i\in I} a_i.\bigoplus_{j\in J_i}p_j \Cdot Q_{ij},\\
& \text{\tiny provided} & \mathord{\inits{R}} \subseteq \sset{a_i}_{i\in I}\\
   \end{array}
  \end{displaymath}

  \caption{Inequations}
  \label{fig:inequations}
\end{figure}

\noindent
In order to characterise the simulation preorders, and the associated
testing preorders, we introduce \emph{inequations}. We write $P \MayAxleq
Q$ when $P\sqsubseteq Q$ is derivable from the inequational theory
obtained by adding the four
\emph{may} inequations in \Fig{fig:inequations} to the
equations in \Fig{fig:cequations}.
The first three additions,
\Ename{May0}--\Ename{May2}, are used in the standard testing theory of
\CSP \cite{csp,DNH84,henn}. For the \emph{must} case, in addition to the
standard inequation \Ename{Must1}, we require an inequational schema,
\Ename{Must2}; this uses the notation $\inits{P}$ to denote the
(finite) set of  initial actions of $P$. Formally,
\[\begin{array}{@{}r@{~=~}l@{}}
\inits{0} & \emptyset\\
\inits{a.P} & \sset{a}\\
\inits{P \probc{p} Q} & \inits{P} \cup \inits{Q}\\
\inits{P \extc Q} & \inits{P} \cup \inits{Q}\\
\inits{P \intc Q} & \sset{\tau}
\end{array}\]
The axiom \Ename{Must2} can equivalently be formulated as follows:
$$\displaystyle \bigoplus_{k\in K}\bigbox_{\ell\in L_k}a_{k\ell}.R_{k\ell}
                \intc \bigsqcap_{i\in I}\bigoplus_{j\in J_i}p_j
                   \Cdot (a_i.Q_{ij} \extc P_{ij}) ~\sqsubseteq~
                   \bigbox_{i\in I} a_i.\bigoplus_{j\in J_i}p_j \Cdot Q_{ij},$$
$$\text{\tiny provided}\qquad
 \{a_{k\ell} \mid k\in K, \ell \in K_k\} \subseteq \{a_i \mid i\in I\}\,.$$
This is the case because a term $R$ satisfies $\mathord{\inits{R}} \subseteq \sset{a_i}_{i\in I}$
iff it can be converted into the form
$\displaystyle\bigoplus_{k\in K}\bigbox_{\ell\in L_k}a_{k\ell}.R_{k\ell}$
by means of axioms \Ename{D1}, \Ename{P1}--\Ename{P3} and
\Ename{E1}--\Ename{E3} of Figure~\ref{fig:inequations}.
This axiom can also be reformulated in an equivalent but more semantic style:
\begin{displaymath}
\begin{array}[t]{@{}lrl@{}}
\Ename{Must2'} & R \intc \bigsqcap_{i\in I}P_i & \sqsubseteq~
                    \bigbox_{i\in I} a_i.Q_i,\\[3pt]
& \text{\tiny provided}~~  \Op{P_i}\ar{a_i}\Op{Q_i}
& \text{\tiny and}~~       \Op{R}\nar{X} \mbox{ with }
                          X=\Act\backslash\sset{a_i}_{i\in I}.

   \end{array}
  \end{displaymath}
This is the case because $\Op{P}\ar{a}\Op{Q}$ iff, up to the axioms
in Figure~\ref{fig:cequations}, $P$ has the form
$\bigoplus_{j\in J}p_j \Cdot (a.Q_{j} \extc P_{j})$
and $Q$ has the form $a.\bigoplus_{j\in J}p_j \Cdot Q_{j}$
for certain $P_j$, $Q_j$ and $p_j$, for $j \inp J$.

Note that \Ename{Must2} can be used, together with \Ename{I1},
to derive the dual of \Ename{May3} via the following inference:
\[\begin{array}{rll}
a.P \probc{p} a.Q & \Eeq & (a.P \probc{p} a.Q)\intc (a.P \probc{p} a.Q)\\
                  & \MustAxleq  & a.(P \probc{p} Q)
\end{array}\]
where we write $P \MustAxleq Q$ when $P\sqsubseteq Q$ is derivable from
the resulting  inequational theory.

An important inequation that follows from \Ename{May1} and \Ename{P1}
is \[\Ename{May4} \quad P \probc{p} Q ~\MayAxleq~ P \intc Q\] saying
that any probabilistic choice can be simulated by an internal choice.
It is derived as follows:
\[\begin{array}{rll}
P \probc{p} Q & \MayAxleq & (P\intc Q)\probc{p} (P\intc Q)\\
              & \Eeq  & (P \intc Q)
\end{array}\]
Likewise, we have  \[\quad P \intc Q ~\MustAxleq~ P \probc{p} Q\;.\]

\begin{thm}\label{thm:complete} For $P,\;Q$ in \nCSP, it holds that
  \begin{enumerate}[\em(i)]
   \item $P \forsimpreo Q$ if and only if $P \MayAxleq Q$
   \item $P \failsimpreo Q$ if and only if $P \MustAxleq Q$.
   \end{enumerate}
\end{thm}
\begin{proof}
  For one direction it is sufficient to check that the inequations,
  and the inequational schema in \Fig{fig:inequations} are
  sound. For $\forsimpreo$ this has been done in \cite{DGHMZ07}, and
  the soundness of \Ename{Must1} and \Ename{Must2'} for $\failsimpreo$
  is trivial. The converse, completeness, is established in the next section.
\end{proof}

%-------------------------------------------------------------------------
\section{Completeness}

\noindent
The completeness proof of \Thm{thm:complete} depends on
the following variation on the \emph{Derivative lemma} of \cite{ccs}:

\begin{lem}[Derivative lemma] Let $P,Q \in \nCSP$.
  \begin{enumerate}[\em(i)]
\item If $\Op{P} \dar{\hat{\tau}} \Op{Q}$ then $P \MustAxleq Q$ and $Q \MayAxleq P$.
\item If  $\Op{P} \dar{a} \Op{Q}$ then $a.Q \MayAxleq P$.
 \end{enumerate}
\end{lem}
\begin{proof}
The proof of (i) proceeds in four stages. We only deal with
$\MayAxleq$, as the proof for $\MustAxleq$ is entirely analogous.

First we show by structural induction on $s\in \sCSP \cap \nCSP$ that
$s \ar{\tau} \Op{Q}$ implies $Q \MayAxleq s$.
So suppose $s \ar{\tau} \Op{Q}$. In case $s$ has the form $P_1 \intc P_2$
it follows by the operational semantics of $\pCSP$ that $Q=P_1$ or $Q=P_2$.
Hence $Q \MayAxleq s$ by $\Ename{May1}$.
The only other possibility is that $s$ has the form $s_1\extc s_2$.
In that case there must be a distribution $\Delta$ such that either
$s_1 \ar{\tau} \Delta$ and $\Op{Q} = \Delta \extc s_2$, or
$s_2 \ar{\tau} \Delta$ and $\Op{Q} = s_1 \extc \Delta$.
Using symmetry, we may restrict attention to the first case.
Let $R$ be a term such that $\Op{R}=\Delta$. Then $\Op{R\extc s_2} =
\Delta \extc s_2 = \Op{Q}$, so \Lem{lem:prob} yields $Q \Peq R\extc s_2$.
By induction we have $R\MayAxleq s_1$, hence $R\extc s_2 \MayAxleq s_1\extc s_2$,
and thus $Q \MayAxleq s$.\vspace{2pt}

Now we show that $s \hatar{\tau} \Op{Q}$ implies $Q \MayAxleq s$.
This follows because $s \hatar{\tau} \Op{Q}$ means that either
$s\ar{\tau}\Op{Q}$ or $\Op{Q}=\pdist{s}$, and in the latter case
\Lem{lem:prob} yields $Q \Peq s$.

Next we show that $\Op{P} \ar{\hat{\tau}} \Op{Q}$ implies $Q \MayAxleq P$.
So suppose  $\Op{P} \ar{\hat{\tau}} \Op{Q}$, that is
  \[\Op{P}=\sum_{i\in I}p_i \cdot \pdist{s_i}
  \qquad s_i \hatar{\tau} \Op{Q_i}
  \qquad \Op{Q}=\sum_{i\in I}p_i \cdot \Op{Q_i}\]
  for some $I$, $p_i \in (0,1]$, $s_i\in\sCSP\cap \nCSP$ and $Q_i\in\nCSP$.
Now
  \begin{enumerate}[(1)]
  \item $\Op{P}=\Op{\bigoplus_{i\in I}p_i \Cdot s_i}$. By
  \Lem{lem:prob} we have $P \Peq \bigoplus_{i\in I}p_i \Cdot s_i$.
  \item $\Op{Q}=\Op{\bigoplus_{i\in I}p_i \Cdot Q_i}$. Again
  \Lem{lem:prob} yields $Q \Peq \bigoplus_{i\in I}p_i \Cdot Q_i$.
  \item $s_i \hatar{\tau} \Op{Q_i}$ implies $Q_i \MayAxleq s_i$.
  Therefore, $\bigoplus_{\!i\in I}p_i \Cdot Q_i \MayAxleq \bigoplus_{\!i\in I}p_i \Cdot s_i$.
  \end{enumerate}
  Combining (1), (2) and (3) we obtain $Q\MayAxleq P$.

Finally, the general case, when  $\Op{P} \ar{\hat{\tau}}^* \Delta$, is now a simple inductive
argument on the length of the derivation.

The proof of (ii) is similar: first we treat the case when $s \ar{a} \Op{Q}$
by structural induction, using \Ename{May2}; then the case  $\Op{P} \ar{a} \Op{Q}$,
exactly as above; and finally use part (i) to derive the general case.
\end{proof}

\noindent
The completeness result now follows from the following two propositions.

\begin{prop}\label{prop:maycomplete}
Let $P$ and $Q$ be in \nCSP. Then $P \forsimpreo Q$ implies $P \MayAxleq Q$.
\end{prop}
\begin{proof}
The proof is by structural induction on $P$ and $Q$, and we may assume that
both $P$ and $Q$ are in normal form because of Proposition~\ref{p:nf}.
So take $P,Q\in\pCSP$ and suppose the claim has been established for all
subterms $P'$ of $P$ and $Q'$ of $Q$, of which at least one of the two
is a strict subterm. We start by proving that if $P\in\sCSP$ then we have
\begin{equation}\label{forsimcomplete}
P \forsim \Op{Q} \quad\mbox{implies}\quad P \MayAxleq Q.
\end{equation}
There are two cases to consider.
\begin{enumerate}[(1)]

\item $P$ has the form $P_1 \intc P_2$. Since $P_i \MayAxleq P$ we
  know $P_i \forsimpreo P \forsimpreo Q$. We use induction to
obtain $P_i \MayAxleq Q$, from which the result follows using \Ename{I1}.

\item $P$ has the form $\bigbox_{i \in I} a_i.P_i$.
If $I$ contains two or more elements then $P$ may also be written as
$\bigsqcap_{i \in I} a_i.P_i$, using \Ename{May0} and \Ename{D2}, and
we may proceed as in case (1) above.
If $I$ is empty, that is $P$ is $\Cnil$, then we can use \Ename{May2}.
So we are left with the possibility that $P$ is $a.P'$.
Thus suppose that $a.P' \forsim \Op{Q}$. We proceed by a case analysis
on the structure of $Q$.
\begin{enumerate}[$\bullet$]
\item $Q$ is $a.Q'$. We know from $a.P' \forsim \Op{a.Q'}$
  that $\Op{P'} \liftforsim \Theta$ for some $\Theta$ with
  $\Op{Q'}\dar{\hat{\tau}}\Theta$, thus $P' \forsimpreo Q'$.
  Therefore, we have $P' \MayAxleq Q'$ by induction. It follows
  that $a.P' \MayAxleq a.Q'$.
\item $Q$ is $\bigbox_{j\in I}a_j.Q_j$ with at least two elements in
  $J$. We use \Ename{May0} and then proceed as in the next case.
\item $Q$ is $Q_1\intc Q_2$. We know from $a.P' \forsim
  \Op{Q_1\intc Q_2}$ that $\Op{P'} \liftforsim \Theta$ for some
  $\Theta$ such that one of the following two conditions holds
  \begin{enumerate}[(a)]
  \item $\Op{Q_i}\dar{a}\Theta$ for $i=1$ or $2$. In this case,
  $a.P' \forsim \Op{Q_i}$, hence $a.P' \forsimpreo Q_i$. By induction
  we have $a.P' \MayAxleq Q_i$; then we apply \Ename{May1}.
 \item $\Op{Q_1}\dar{a}\Theta_1$ and $\Op{Q_2}\dar{a}\Theta_2$ such
  that $\Theta=p\cdot\Theta_1 + (1-p)\cdot \Theta_2$ for some $p\in (0,1)$.
  Let $\Theta_i = \Op{Q'_i}$ for $i=1,2$. By the Derivative Lemma, we have
  $a.Q'_1 \MayAxleq Q_1$ and $a.Q'_2 \MayAxleq Q_2$. Clearly,
  $\Op{Q'_1 \probc{p} Q'_2}=\Theta$, thus $P' \forsimpreo Q'_1 \probc{p} Q'_2$.
  By induction, we infer that $P'\MayAxleq Q'_1 \probc{p} Q'_2$. So
  $$\begin{array}{rl@{\qquad}l}
  a.P' & \MayAxleq~ a.(Q'_1 \probc{p} Q'_2)\\
       & \MayAxleq~ a.Q'_1 \probc{p} a.Q'_2 &{\Ename{May3}}\\
       & \MayAxleq~ Q_1 \probc{p} Q_2\\
       & \MayAxleq~ Q_1 \intc Q_2 & {\Ename{May4}}
  \end{array}$$
  \end{enumerate}
\item $Q$ is $Q_1 \probc{p} Q_2$. We know from $a.P' \forsim \Op{Q_1
  \probc{p} Q_2}$ that $\Op{P'} \liftforsim \Theta$ for some
  $\Theta$ such that $\Op{Q_1 \probc{p} Q_2}\dar{a}\Theta$.
  From Lemma~\ref{lem:sum} we know that $\Theta$ must take the form
  $p\cdot\Op{Q'_1}+(1-p)\cdot\Op{Q'_2}$, where $\Op{Q_i}\dar{a}\Op{Q'_i}$ for
  $i=1,2$. Hence $P' \forsimpreo Q'_1 \probc{p} Q'_2$, and by
  induction we get $P' \MayAxleq Q'_1 \probc{p} Q'_2$. Then we can
  derive $a.P' \MayAxleq Q_1 \probc{p} Q_2$ as in the previous case.
\end{enumerate}
\end{enumerate}
Now we use \Eqn{forsimcomplete} to show that $P \forsimpreo Q$ implies $P \MayAxleq Q$.
Suppose $P \forsimpreo Q$. Applying \Def{def:simulation preorder}
with the understanding that any distribution $\Theta\in\dist{\sCSP}$
can be written as $\Op{Q'}$ for some $Q'\inp\pCSP$, this means that
$\Op{P} \liftforsim \Op{Q'}$ for some $\Op{Q}\dar{\hat{\tau}}\Op{Q'}$.\linebreak
The Derivative Lemma yields $Q' \MayAxleq Q$. So it suffices to show $P \MayAxleq Q'$.
We know that $\Op{P} \liftforsim \Op{Q'}$ means that
  \[\Op{P}=\sum_{k\in K}r_k \cdot \pdist{t_k}
  \qquad t_k \forsim \Op{Q'_k}
  \qquad \Op{Q'}=\sum_{k\in K}r_k \cdot \Op{Q'_k}\]
  for some $K$, $r_k \in (0,1]$, $t_k\in\sCSP$ and $Q'_k\in\pCSP$.
Now
  \begin{enumerate}[(1)]
  \item $\Op{P}=\Op{\bigoplus_{k\in K}r_k \Cdot t_k}$. By
  \Lem{lem:prob} we have $P \Peq \bigoplus_{k\in K}r_k \Cdot t_k$.
  \item $\Op{Q'}=\Op{\bigoplus_{k\in K}r_k \Cdot Q'_k}$. Again
  \Lem{lem:prob} yields $Q' \Peq \bigoplus_{k\in K}r_k \Cdot Q'_k$.
  \item $t_k \forsim \Op{Q'_k}$ implies $t_k \MayAxleq\! Q'_k$ by \Eqn{forsimcomplete}.
  Therefore, $\bigoplus_{\!k\in K}r_k \Cdot t_k \MayAxleq\! \bigoplus_{\!k\in K}r_k \Cdot Q'_k$.
  \end{enumerate}
  Combining (1), (2) and (3) we obtain $P\MayAxleq Q'$, hence $P\MayAxleq Q$.
\end{proof}

\begin{prop}\label{prop:mustcomplete}
Let $P$ and $Q$ be in \nCSP. Then $P \failsimpreo Q$ implies $P \MustAxleq Q$.
\end{prop}
\begin{proof}
Similar to the proof of
\Prop{prop:maycomplete}, but using a reversed orientation of the
preorders. The only real difference is the case (2), which we consider now.
So assume $Q \failsim \Op{P}$, where $Q$ has the form $\bigbox_{i \in I} a_i.Q_i$.
Let $X$ be any set of actions such that $X\cap\sset{a_i}_{i\in I}=\emptyset$;
then $\bigbox_{i\in I}a_i.Q_i \nar{X}$. Therefore, there exists a $P'$ such that
$\Op{P}\dar{\hat{\tau}}\Op{P'}\nar{X}$. By the Derivative lemma,\vspace{-5pt}
\begin{equation}\label{e:ref}
P\MustAxleq P'\vspace{5pt}
\end{equation}
Since $\bigbox_{i\in I}a_i.Q_i\ar{a_i}\Op{Q_i}$, there exist
$P_i,P'_i,P''_i$ such that
$\Op{P}\dar{\hat{\tau}}\Op{Pi}\ar{a_i}\Op{P'_i}\dar{\hat{\tau}}\Op{P''_i}$ and
$\Op{Q_i}\liftfailsim \Op{P''_i}$. Now\vspace{-5pt}
\begin{equation}\label{e:act}
P\MustAxleq P_i\vspace{5pt}
\end{equation}
using the Derivative lemma, and $P'_i \failsimpreo Q_i$, by \Def{def:simulation preorder}.
By induction, we have $P'_i \MustAxleq Q_i$, hence\vspace{-5pt}
\begin{equation}\label{e:actm}
\bigbox_{i\in I}a_i.P'_i \MustAxleq \bigbox_{i\in I} a_i.Q_i\vspace{5pt}
\end{equation}
The desired result is now obtained as follows:\\[1ex]
\mbox{~}\hfill$\begin{array}[b]{cl@{\quad}l}\displaystyle
P &\displaystyle \MustAxleq~ P'\intc \bigsqcap_{i\in I}P_i &
                          \mbox{by \Ename{I1}, (\ref{e:ref}) and (\ref{e:act})}\\
  &\displaystyle \MustAxleq~ \bigbox_{i\in I}a_i.P'_i & \mbox{by \Ename{Must2'}}\\
  &\displaystyle \MustAxleq~ \bigbox_{i\in I}a_i.Q_i  & \mbox{by (\ref{e:actm})}
\end{array}$
\end{proof}
\noindent
Propositions~\ref{prop:maycomplete} and~\ref{prop:mustcomplete} give
us the completeness result stated in Theorem~\ref{thm:complete}.

%-------------------------------------------------------------------------
\section{Conclusions and related work}\label{sec:conc}

\noindent
In this paper we continued our previous work \cite{DGHMZ07,DGMZ07}
in our quest for a testing theory for processes which exhibit both
nondeterministic and probabilistic behaviour. We have studied
three different aspects of may- and must testing preorders for finite processes: (i) we have
shown that the may preorder can be characterised as a co-inductive
simulation relation, and the must preorder as a failure simulation
relation; (ii) we have given a characterisation of both preorders in a
finitary modal logic; and (iii) we have also provided complete
axiomatisations for both preorders over a probabilistic version of
recursion-free CSP. Although we omitted our parallel operator
$\Par{A}$ from the axiomatisations, it
and similar CSP and CCS-like parallel operators can be handled using standard
techniques, in the must case at the expense of introducing auxiliary
operators. In future work we hope to extend these results to recursive processes.

We believe these results, in each of the three areas, to be novel, although
a number of partial results along similar lines exist in the literature.
These are detailed below.

\paragraph{\bf Related work:}
Early additions of probability to CSP include work by Lowe
\cite{Low93}, Seidel \cite{Sei95} and Morgan et al.\ \cite{MMSS96};
but all of them were forced to make compromises of some kind in
order to address the potentially complicated interactions between
the three forms of choice. The last \cite{MMSS96} for example
applied the Jones/Plotkin probabilistic powerdomain \cite{Jones89}
directly to the failures model of CSP \cite{BHR84}, the resulting
compromise being that probability distributed outwards
through all other operators; one controversial result of that was
that internal choice was no longer idempotent, and that it was
``clairvoyant'' in the sense that it could adapt to
probabilistic-choice outcomes that had not yet occurred.
Mislove addressed this problem in \cite{Mis00} by presenting a
denotational model in which internal choice distributed outwards
through probabilistic choice.
However, the distributivities of both \cite{MMSS96} and \cite{Mis00} constitute
identifications that cannot be justified by our testing approach; see \cite{DGHMZ07}.

In Jou and Smolka \cite{JS90}, as in \cite{Low93,Sei95}, probabilistic
equivalences based on traces, failures and readies are defined. These
equivalences are coarser than $\pMayeq$.  For example, the two
processes in Example~\ref{ex:1} cannot be distinguished by the
equivalences of \cite{JS90,Low93,Sei95}.  However, we can tell them
apart by the test given in Example~\ref{ex:2}.

Probabilistic extensions of testing equivalences \cite{DNH84} have
been widely studied.
There are two different proposals on how to include probabilistic choice:
  (i) a test should be non-probabilistic, that is there is no occurrence
  of probabilistic choice in a test
  \cite{LS91,Chr90,JHW94,KN98,GN99}; or (ii) a test can be
  probabilistic, that is probabilistic choice may occur in tests as well
  as processes \cite{CDSY99,WL92,Nun03,JW95,Seg96,JW02,CCVP03}.
  This paper adopts the second approach.

Some work \cite{LS91,Chr90,CDSY99,Nun03} does not consider nondeterminism
  but deals exclusively with \emph{fully probabilistic} processes. In this setting a process
  passes a test with a unique probability instead of a set of probabilities,
  and testing preorders in the style of \cite{DNH84} have been
  characterised in terms of \emph{probabilistic traces}
  \cite{CDSY99} and \emph{probabilistic acceptance trees} \cite{Nun03}.
Cazorla et al.\ \cite{CCVP03} extended the results of \cite{Nun03}
with nondeterminism, but suffered from the same problems as \cite{MMSS96}.

The work most closely related to ours is \cite{JW95,JW02}.
  In \cite{JW95} Jonsson and Wang characterised may- and must-testing
  preorders in terms of ``chains'' of traces and failures,
  respectively, and in \cite{JW02} they presented a ``substantially
  improved'' characterisation of their may-testing preorder
  using a notion of simulation which is weaker than $\forsimpreo$
(cf.\ \Def{def:simulation preorder}).
They only considered processes without $\tau$-moves. In
 \cite{DGHMZ07} we have shown that tests
with internal moves can distinguish more processes than tests without
internal moves, even when applied to processes that have no internal moves
themselves.

Segala \cite{Seg96} defined two preorders called trace
distribution precongruence ($\tdc$) and failure distribution
precongruence ($\fdc$). He proved that the former coincides with an
infinitary version of $\pdmMayleq$ (cf.\ \Def{def:vector-testing})
and that the latter coincides with an infinitary version of \plat{$\pdmMustleq$}.
In \cite{LSV03} it has been shown that $\tdc$
coincides with a notion of simulation akin to $\forsimpreo$.
Other probabilistic extensions of simulation occurring in the
literature are reviewed in \cite{DGHMZ07}.

%------------------------------------------------
\appendix

\section{Resolution-based testing}
\label{app:resolutions}

\noindent
A \emph{probabilistic automaton} consists of a pLTS $\langle S, L,
\rightarrow \rangle$ and a distribution $\DeltaC$ over $S$. Since we
only consider probabilistic automata with $L=\Act_\tau \cup\Omega$, we
omit it and write a probabilistic automaton simply as a triple
$\langle S,\DeltaC,\rightarrow\rangle$ and call $\DeltaC$ the
\emph{initial distribution} of the automaton. The operational
semantics of a \plat{\pCSPO} process $P$ can thus be viewed as a
probabilistic automaton with initial distribution $\DeltaC :=
\Op{P}$. States in a probabilistic automata that are not reachable
from the initial distribution are generally considered irrelevant and
can be omitted.

A probabilistic automaton is called \emph{finite} if there exists a
function $\depth: S \cup \dist{S} \rightarrow \mbox{\bbb N}$ such that
$s \in \support{\Delta}$ implies $\depth(s) < \depth(\Delta)$ and
$s \ar{\alpha} \Delta$ implies $\depth(s) > \depth(\Delta)$.
Finite probabilistic automata can be drawn as explained at the end of
Section~\ref{sec:pcsp}.

A \emph{fully probabilistic automaton} is one in which each state
enables at most one action, and (general) probabilistic automata can
be ``resolved'' into fully probabilistic automata by pruning away
multiple action-choices until only single choices are left, possibly
introducing some linear combinations in the process. We define this
formally for probabilistic automata representing \pCSPO expressions.
\begin{defi}\cite{DGMZ07}\label{d:resolution}
A \emph{resolution} of a distribution \plat{$\DeltaC \in \dist{\sCSPO}$}
is a fully probabilistic automaton $\langle R,\Theta^\circ,\rightarrow\rangle$ such
that there is a resolving function $f: R\rightarrow \sCSPO$ which satisfies:
\begin{enumerate}[(i)]
\item $f(\Theta^\circ)=\DeltaC$
\item if $r\ar{\alpha}\Theta$ then $f(r)\ar{\alpha}f(\Theta)$
\item if $r\not\rightarrow$ then $f(r)\not\rightarrow$
\end{enumerate}
where $f(\Theta)$ is the distribution defined by
$f(\Theta)(s):=\sum_{f(r)=s}\Theta(r)$.
\end{defi}

\noindent
Note that resolutions of distributions $\DeltaC\in\dist{\sCSPO}$ are
always finite.
We define a function which yields the probability that a given fully
probabilistic automaton will start with a particular sequence of actions.
\begin{defi}\cite{DGMZ07}
Given a fully probabilistic automaton $\eR=\langle R,\DeltaC,\rightarrow\rangle$,
the probability that $\eR$ follows the sequence of actions $\sigma\in\Sigma^*$
from its initial distribution is given by $\PrR(\sigma,\DeltaC)$, where
$\PrR:\Sigma^*\times R\rightarrow[0,1]$ is defined inductively by
\[\PrR(\varepsilon,r):=1 \quad\mbox{ and }\quad
\PrR(\alpha\sigma,r):=\left\{\begin{array}{ll}
\PrR(\sigma,\Delta) & \mbox{if $r\ar{\alpha}\Delta$}\\
0 & \mbox{otherwise}
\end{array}\right.\]
and
$\PrR(\sigma,\Delta):=\Exp{\Delta}(\PrR(\sigma,\underline{~\;}))=
\sum_{r\in\support{\Delta}}\Delta(r)\cdot\PrR(\sigma,r)$.
Here $\varepsilon$ denotes the empty sequence of actions and
$\alpha\sigma$ the sequence starting with $\alpha\in\Sigma$
and continuing with $\sigma\in\Sigma^*$. The value $\PrR(\sigma,r)$ is the probability
that $\eR$ proceeds with sequence $\sigma$ from state $r$.

Now let $\Sigma^{*\alpha}$ be the set of finite sequences in $\Sigma^*$ that
contain $\alpha$ exactly once, and that at the end. Then the probability
that the fully probabilistic automaton $\eR$ ever performs an action
$\alpha$ is given by
$\sum_{\sigma\in\Sigma^{*\alpha}}\PrR(\sigma,\DeltaC)$.
\end{defi}

\noindent
We recall the results-gathering function ${\mathbb W}$ given in
Definition 5 of \cite{DGMZ07}.

\begin{defi}\label{def:results}
For a fully probabilistic automaton $\eR$, let its \emph{success tuple}
$\WW(\eR)\in[0,1]^\Omega$ be such that $\WW(\eR)(\omega)$ is the probability that $\eR$
ever performs the action $\omega$.

Then for a distribution $\DeltaC\in\dist{\sCSPO}$ we
define the \emph{set} of its success tuples to be those resulting as above
from all its resolutions separately:
\[\WW(\DeltaC) ~:=~\sset{\WW(\eR)\mid \eR \mbox{ is a resolution of }\DeltaC}.\]
\end{defi}

\noindent
We relate these sets of tuples to Definition~\ref{def:vector-testing},
in which similar sets are produced ``all at once,'' that is without
introducing resolutions first. In fact we will find that they are the
same. Note that Definition~\ref{def:vector-testing} of $\MDCresultsN$
extends smoothly to states and distributions in probabilistic
automata.  When applied to fully probabilistic automata,
\plat{$\MDCresultsN$} always yields singleton sets, which we will
loosely identify with their unique members; thus when we write
\plat{$\MDCresults{\Delta}(\omega)$} with $\Delta$ a distribution in a
fully probabilistic automaton, we actually mean the $\omega$-component
of the unique element of \plat{$\MDCresults{\Delta}$}.

\begin{lem}\label{lem:full.occ}
If $\eR=\langle R,\DeltaC,\rightarrow \rangle$ is a finite fully
probabilistic automaton, then
\begin{enumerate}[(1)]
\item $\MDresults{\Delta}=\MDCresults{\Delta}$ for all $\Delta\in \dist{R}$, and
\item ${\mathbb W}(\eR)=\MDresults{\DeltaC}$.
\end{enumerate}
\end{lem}

\noindent\textit{Proof.\/}
(1) is immediate: since the automaton is fully probabilistic, convex closure has no
effect.  For (2) we need to show that for all $\omega\in\Omega$ we have
$\WW(\eR)(\omega)=\MDresults{\DeltaC}(\omega)$, i.e.\ that
$\sum_{\sigma\in\Sigma^{*\omega}}\PrR(\sigma,\DeltaC)=(\MDresults{\DeltaC})(\omega)$.
So let $\omega\in\Omega$.  We show
\begin{equation}
\sum_{\sigma\in\Sigma^{*\omega}}\PrR(\sigma,\Delta)=\MDresults{\Delta}(\omega)
\qquad \mbox{and} \qquad
\sum_{\sigma\in\Sigma^{*\omega}}\PrR(\sigma,r)=\MDresults{r}(\omega)
\end{equation}
for all $\Delta \in \dist{R}$ and $r\in R$, by simultaneous induction on the depths of
$\Delta$ and $r$.
\begin{enumerate}[$\bullet$]
\item In the base case $r$ has no enabled actions. Then
  $\forall i:\sum_{\sigma\in\Sigma^{*\omega}}\PrR(\sigma,r)=0$ and
  $\MDresults{r}=\vec{0}$, so $\MDresults{r}(\omega)=0$.
\item Now suppose there is a transition
  $r\ar{\alpha}\Delta$ for some action $\alpha$ and distribution
  $\Delta$. There are two possibilities:
\begin{enumerate}[$-$]
\item $\alpha=\omega$. We then have $\MDresults{s}(\omega)=1$. Now for any
  finite non-empty sequence $\sigma$ without any occurrence of $\omega$ we have
  $\PrR(\sigma\omega,r)=0$. Thus
  $\sum_{\sigma\in\Sigma^{*\omega}}\PrR(\sigma,r)= \PrR(\omega,r)=1$ as required.
\item $\alpha\not=\omega$. Since
  $\MDresults{r}=\alpha!\MDresults{\Delta}$, we have
  $\MDresults{r}(\omega)=\MDresults{\Delta}(\omega)$. On the other hand,
  $\PrR(\beta\sigma,r)=0$ for $\beta\not=\alpha$. Therefore
  \[\begin{array}{rcl}
  \sum_{\sigma\in\Sigma^{*\omega}}\PrR(\sigma,r)
  & = & \sum_{\alpha\sigma\in\Sigma^{*\omega}}\PrR(\alpha\sigma,r)\\
  & = & \sum_{\sigma\in\Sigma^{*\omega}}\PrR(\alpha\sigma,r)\\
  & = & \sum_{\sigma\in\Sigma^{*\omega}}\PrR(\sigma,\Delta)\\
  & = & \MDresults{\Delta}(\omega) \qquad\mbox{by induction}\\
  & = & \MDresults{r}(\omega)~.
  \end{array}\]
\end{enumerate}
\item
Finally, $\sum_{\sigma\in\Sigma^{*\omega}}\PrR(\sigma,\Delta)=
          \sum_{\sigma\in\Sigma^{*\omega}}\Exp\Delta(\PrR(\sigma,\underline{~\;}))=
          \Exp{\Delta}(\sum_{\sigma\in\Sigma^{*\omega}}\PrR(\sigma,\underline{~\;}))\\
=         \Exp{\Delta}(\MDresults{\underline{~\;}}(\omega))=
          \Exp{\Delta}(\MDresults{\underline{~\;}})(\omega)=
          \MDresults{\Delta}(\omega)$.
\hfill\qed
\end{enumerate}

\noindent
Now we look more closely at the interaction of $\MDCresultsN$ and resolutions.

\begin{lem}\label{lem:reso.occ}
Let $\DeltaC\in\dist{\sCSPO}$.
\begin{enumerate}[(1)]
\item If $\langle R,\Theta^\circ,\rightarrow\rangle$ is a resolution of $\DeltaC$, then
  $\MDCresults{\Theta^\circ}\in \MDCresults{\DeltaC}$.
\item If $o\in\MDCresults{\DeltaC}$ then there is a resolution
  $\langle R,\Theta^\circ,\rightarrow\rangle$ of $\DeltaC$ such that
  $\MDCresults{\Theta^\circ}=o$.
\end{enumerate}
\end{lem}
\noindent\textit{Proof.\/}
\begin{enumerate}[(1)]
\item Let $\langle R,\Theta^\circ,\rightarrow\rangle$ be a resolution
  of $\DeltaC$ with resolving function $f$. We observe that for any
  $\Theta \in\dist{R}$ we have
  \begin{equation}\label{eq:obs}
  \forall r\in \support{\Theta}:\MDCresults{r}\in\MDCresults{f(r)}
  \text{\ implies\ }  \MDCresults{\Theta}\in\MDCresults{f(\Theta)}
  \end{equation}
  because
  \[\begin{array}{rcl}
  \MDCresults{\Theta} & = &
  \sum_{r\in\support{\Theta}}\Theta(r)\cdot\MDCresults{r} \\
  & \in &\sum_{r\in\support{\Theta}}\Theta(r)\cdot\MDCresults{f(r)} \\
  & = & \sum_{s\in\support{f(\Theta)}}f(\Theta)(s)\cdot\MDCresults{s}\\
  & = & \MDCresults{f(\Theta)}~.
  \end{array}\]
  We now prove by induction on $\depth(r)$ that $\forall r\in T:
  \MDCresults{r}\in \MDCresults{f(r)}$, from which the required result
  follows in view of (\ref{eq:obs}) and the fact that
  $f(\Theta^\circ)=\DeltaC$.
  \begin{enumerate}[$\bullet$]
  \item In the base case we have $r\not\rightarrow$, which implies
  $f(r)\not\rightarrow$. Therefore, we have
  $\MDCresults{r}=\vec{0}\in\MDCresults{f(r)}$.
  \item Otherwise $r$ has a transition
  $r\ar{\alpha}\Theta$ for some $\alpha$ and $\Theta$.  By induction
  we have $\MDCresults{r'}\in\MDCresults{f(r')}$ for all
  $r'\in\support{\Theta}$. Using (\ref{eq:obs}) we get
  $\MDCresults{\Theta}\in\MDCresults{f(\Theta)}$. Now
  \[\MDCresults{r}~=~\alpha!\MDCresults{\Theta} ~\in~
  \alpha!\MDCresults{f(\Theta)}~\subseteq~ \MDCresults{f(r)}\] where the
  last step follows from the fact that $f(r)\ar{\alpha}f(\Theta)$ is
  one of the transitions of $f(r)$.
  \end{enumerate}
\item This clause is proved by induction on $\depth(\DeltaC)$.
  First consider the special case that $\DeltaC$ is a point
  distribution on some state $s$.
  \begin{enumerate}[$\bullet$]
  \item In the base case we have $s\not\rightarrow$. The probabilistic automaton
  $\langle \{s\},\pdist{s},\emptyset\rangle$ is a resolution of
  $\DeltaC=\pdist{s}$ with the resolving function being the identity.
  Clearly, this resolution satisfies our  requirement.
  \item Otherwise there is a finite, non-empty index set $I$ such that
  $s\ar{\alpha_i}\Delta_i$ for some actions $\alpha_i$ and
  distributions $\Delta_i$. If
  \plat{$o\in\MDCresults{\DeltaC}=\MDCresults{s}$}, then by the definition
  of \plat{$\MDCresultsN$} we have
  $o=\sum_{i\in I}p_i\Cdot \alpha_i!o_i$ with
  \plat{$o_i\in\MDCresults{\Delta_i}$} and $\sum_{i\in I}p_i=1$ for some
  $p_i\in[0,1]$. By induction, for each $i\in I$ there is a resolution
  $\langle R_i,\Theta_i^\circ,\rightarrow_i\rangle$ of $\Delta_i$ with
  resolving function $f_i$ such that \plat{$\MDCresults{\Theta_i^\circ}=o_i$}.
  Without loss of generality, we assume that $R_i$ is disjoint from
  $R_j$ for $i\not=j$, as well as from $\{r_i \mid i\in I\}$. We now
  construct a fully probabilistic automaton
  $\langle R,\Theta^\circ,\rightarrow'\rangle$ as follows:
  \begin{enumerate}[$-$]
  \item[$\bullet$] $R:=\sset{r_i\mid i\in I}\cup\bigcup_{i\in I}R_i$
  \item[$\bullet$] $\Theta^\circ:=\sum_{i\in I}p_i\cdot\pdist{r_i}$
  \item[$\bullet$] $\rightarrow':=\sset{r_i\ar{\alpha_i}\Theta_i^\circ\mid i\in
  I}\cup\bigcup_{i\in I} \rightarrow_i$.
  \end{enumerate}
  This automaton is a resolution of $\DeltaC = \pdist{s}$ with resolving function $f$
  defined by
  \[f(r)=\left\{\begin{array}{ll}
  s & \mbox{if $r=r_i$ for $i\in I$}\\
  f_i(r) & \mbox{if $r\in R_i$ for $i\in I$.}
  \end{array}\right.\]
  The resolution thus constructed satisfies our requirement because
  \[\begin{array}{rcl}
  \MDCresults{\Theta^\circ} & = & \MDCresults{\sum_{i\in
  I}p_i\cdot\pdist{r_i}}\\
  & = & \sum_{i\in I}p_i\cdot\MDCresults{r_i}\\
  & = & \sum_{i\in I}p_i\cdot\alpha_i!\MDCresults{\Theta_i^\circ}\\
  & = & \sum_{i\in I}p_i\cdot\alpha_i! o_i\\
  & = & o ~.
  \end{array}\]
  \end{enumerate}
  We now consider the general case that $\DeltaC$ is a proper
  distribution with $\support{\DeltaC} = \sset{s_j\!\mid j\inp J}$ for
  some finite index set $J$. Using the reasoning in the above special
  case, we have a resolution $\langle R_j, \Theta_j^\circ,
  \rightarrow_j\rangle$ of each distribution $\pdist{s_j}$.  Without
  loss of generality, we assume that $R_j$ is disjoint from $R_{k}$
  for $j\not=k$. Consider the probabilistic automaton $\langle
  \bigcup_{j\in J}R_j,\sum_{j\in J}\DeltaC(s_j)\cdot\Theta_j^\circ,
  \bigcup_{j\in J}\rightarrow_j\rangle$. It is a resolution of
  $\DeltaC$ satisfying our requirement. If $o\in\MDCresults{\DeltaC}$
  then $o=\sum_{j\in J}\DeltaC(s_j)\Cdot o_j$ with $o_j\in\MDCresults{s_j}$.
  Since $o_j=\MDCresults{\Theta_j^\circ}$, we have
  \plat{$o=\MDCresults{\sum_{j\in J}\DeltaC(s_j)\cdot\Theta_j^\circ}$}.
  \hfill\qed
\end{enumerate}

\noindent
We can now give the result relied on in Section~\ref{sec:vector}.

\begin{prop}\label{p1126}
Let $\DeltaC\in\dist{\sCSPO}$. Then we have that
${\mathbb W}(\DeltaC)=\MDCresults{\DeltaC}$.
\end{prop}
\begin{proof}
Combine Lemmas~\ref{lem:full.occ} and \ref{lem:reso.occ}.
\end{proof}

%------------------------------------------------

\bibliographystyle{plain}

\end{document}